\documentclass[final,5p,times,twocolumn]{elsarticle}

\usepackage{amssymb}
\usepackage{amsmath}
\usepackage{amsthm}
\usepackage{bm}

\usepackage{lineno}

\usepackage{booktabs}
\usepackage[colorlinks=true, linkcolor=blue, citecolor=blue, urlcolor=blue]{hyperref}  % Includes url handling
\usepackage{xurl}   % improves URL line breaking

\begin{document}

\begin{frontmatter}

\title{Benchmarking of Massively Parallel Phase-Field Codes for Directional Solidification}

\author[label1]{Jiefu Tian}
\author[label2]{David Montiel}
\author[label1,label3]{Kaihua Ji}
\author[label1]{Trevor Lyons}
\author[label2]{Jason Landini}
\author[label2,label4]{Katsuyo Thornton}
\author[label1]{Alain Karma\corref{cor1}}
\ead{a.karma@northeastern.edu}

\cortext[cor1]{Corresponding author}

\affiliation[label1]{organization={Department of Physics and Center for Interdisciplinary Research on Complex Systems, Northeastern University},
            city={Boston},
            postcode={02115},
            state={MA},
            country={USA}}

\affiliation[label2]{organization={Department of Materials Science and Engineering, University of Michigan},
            city={Ann Arbor},
            postcode={48109},
            state={MI},
            country={USA}}

\affiliation[label3]{organization={Materials Science Division, Lawrence Livermore National Laboratory},
            city={Livermore},
            postcode={94550},
            state={CA},
            country={USA}}

\affiliation[label4]{organization={Department of Nuclear Engineering and Radiological Sciences, University of Michigan},
city={Ann Arbor},
postcode={48109},
state={MI},
country={USA}}

%% Abstract
\begin{abstract}
We present a detailed benchmark comparing two state-of-the-art phase-field implementations for simulating alloy solidification under experimentally relevant conditions. The study investigates the directional solidification of Al-3wt\%Cu under high-velocity solidification conditions and SCN-0.46wt\% camphor under microgravity conditions from National Aeronautics and Space Administration (NASA) DECLIC-DSI-R experiments.
Both codes, one employing finite-difference discretization with uniform mesh and GPU-acceleration (GPU-PF) and the other one employing finite-element discretization with adaptive-mesh and CPU-parallelization (PRISMS-PF), solve the same quantitative phase-field formulation that incorporates an anti-trapping current for the solidification of dilute alloys. We evaluate the predictions of each code for dendritic morphology, primary spacing, and tip dynamics in both 2D and 3D, as well as their numerical convergence and computational performance.
While existing benchmark problems have primarily focused on simplified or small-scale simulations, they do not reflect the computational and modeling challenges posed by employing experimentally relevant time and length scales. Our results provide a practical framework for assessing phase-field code performance as well as validating and facilitating their application in integrated computational materials engineering (ICME) workflows that require integration with realistic experimental data.
\end{abstract}


\begin{highlights}
\item Benchmarked GPU and CPU phase-field codes in 2D and 3D alloy solidification
\item Used an identical model setup to compare finite difference and adaptive FEM implementations
\item Validated against NASA microgravity data for dendrite spacing and morphology
\item Demonstrated convergence and tip-scale agreement across spatial resolutions
\item Compared performance scaling under realistic ICME-relevant simulation conditions
\end{highlights}

%% Keywords

\end{frontmatter}

% \linenumbers


\section{Introduction}
\label{sec:intro}
%% Labels are used to cross-reference an item using \ref command.
Integrated Computational Materials Engineering (ICME) provides a unifying framework for linking materials properties across multiple length and time scales, accelerating materials discovery and design through predictive simulations and data integration~\cite{RN220}. At its core, ICME enables the rigorous integration of physics-based models across multiple scales, from quantum and atomistic to mesoscale and macroscale, within engineering workflows for real-world applications. ICME has demonstrated significant impact in the design and manufacturing of aerospace materials~\cite{RN172}, advanced batteries~\cite{Lvovich2018}, and automotive components~\cite{RN173,RN174}.

At the atomic scale, methods such as Density Functional Theory (DFT) and Molecular Dynamics (MD) have been widely used to understand material properties and behavior at the small length scale and short timescale associated with atomic motion. At the macroscale, continuum mechanics and finite element methods offer predictions of material properties, such as stress, deformation, and thermal gradients in components. Between these scales is the mesoscale, where interfaces in materials, such as those defining grains and dendrites, evolve dynamically, ultimately influencing material behaviors~\cite{RN166}. 

Two general mesoscale modeling approaches exist for treating interfaces: sharp-interface and diffuse-interface methods. Sharp-interface methods assume a discontinuous transition across a zero-thickness interface and operate at the intrinsic physical length and time scales. While suitable for simple geometries, sharp-interface methods require complex numerical techniques to handle topological changes, and evolving complex interfacial structures are often challenging to describe. In contrast, the interfaces in diffuse-interface methods are represented using continuous fields, have a finite thickness $W_0$, and their evolution is governed by variational principles posed in the form of partial differential equations. Thus, diffuse-interface methods enable a robust and seamless treatment of interface evolution and topological transitions. 
%add_revision
Validating both types of implementations under realistic conditions is essential for their adoption in ICME workflows~\cite{NASA_TM_2025}.
%add_revision

The phase-field (PF) method~\cite{RN91,Fix1983} is a widely used diffuse-interface technique. It replaces explicit interface tracking with the evolution of one or more scalar fields $\phi(\mathbf{r}, t)$, which represent phenomenological descriptors of the material phase (or state) at spatial location $\mathbf{r}$ and time $t$. These fields are referred to as phase fields. The evolution equations for these phase fields are typically derived variationally from a Lyapunov functional that represents the total free energy of the system. PF models inherently capture topological transitions, such as branching, merging, and pinching, without the need for ad hoc treatments, making them ideally suited for simulating time-dependent free-boundary problems (TDFBPs) in complex microstructures. Early applications of the PF method included modeling the solidification of pure materials~\cite{RN91,RN164} and spinodal decomposition via the Cahn-Hilliard equation~\cite{RN165}. To date, the PF method has been applied to a much wider range of materials science phenomena. For comprehensive reviews of PF modeling, see Refs~\cite{RN166,RN181,RN180,RN182,RN186,boettinger2010phasefield,RN167}.

Despite the advantages of the PF method, simulations at experimentally relevant length and time scales, approximately on the order of millimeters and seconds, remain a barrier to scientific discoveries due to computational cost. Most implementations operate in the nanometer or a few hundred micrometers range and require massive parallelism for scale-up. As a result, many researchers develop customized, high-performance PF codes tailored to specific physics such as fracture mechanics~\cite{RN2,RN14}, alloy solidification~\cite{RN32,RN163}, crystal nucleation~\cite{RN177}, or surface growth~\cite{RN178,RN179}. These domain-specific codes often use hard-coded solvers to minimize overhead, trading flexibility for efficiency, which complicates long-term maintenance and integration of newer numerical methods. Recently, several modular open-source or commercial PF frameworks have emerged and begun building user communities, including the open-source frameworks PRISMS-PF~\cite{RN186,RN98}, MOOSE~\cite{giudicelli2024moose,RN184,RN185}, MEMPHIS~\cite{osti_1729722}, FiPy~\cite{RN189,RN190}, OpenPhase~\cite{RN191,RN192}, and the commercial software MICRESS~\cite{RN187,RN188}. However, comprehensive benchmarking of these frameworks under physically relevant, experimentally validated conditions remains rare. Most existing benchmarks focus on idealized or small-scale cases~\cite{wheeler2019pfhub,RN168}, which limits the assessment of the scalability and predictive capability of different codes. Furthermore, direct comparisons between flexible modular frameworks and performance-optimized codes remain limited. Validating both types of implementations under realistic conditions is essential for their adoption in ICME workflows.

We choose directional solidification of transparent binary alloys in microgravity as the platform for quantitative benchmarking of PF models. For validation, we leverage the data from the National Aeronautics and Space Administration (NASA) DECLIC-DSI-R (Device for the Study of Critical Liquids and Crystallization - Directional Solidification Insert - Reflight) experiments conducted aboard the International Space Station (ISS)~\cite{RN163,RN111,RN43}. These experiments examined the growth dynamics of a transparent organic alloy of succinonitrile-camphor (SCN-Camphor) at varying concentrations under purely diffusive conditions. In these experiments, directional solidification proceeds with a low pulling velocity $V_p$ and thermal gradient $G$, producing quasi-steady-state dendritic or cellular arrays with primary spacing on the order of a few hundred micrometers. These experiments provide spatial and temporal resolution sufficient to validate tip dynamics, spacing selection, and morphological evolution predicted by quantitative PF models~\cite{RN163,RN43}. The absence of buoyancy-driven convection in microgravity enables purely solutal diffusive solidification, precisely the regime where PF models are most computationally tractable and predictive.

Benchmarking under a unified model framework is critical to enable rigorous, direct (``apples-to-apples'') comparison of accuracy, stability, and computational performance across codes. Notably, two major community platforms use different formulations for alloy solidification applications. PRISMS-PF adopts the thin-interface formulation~\cite{karma2001phase,RN32}, while MOOSE employs the Kim-Kim-Suzuki (KKS)~\cite{RN92} and grand potential models~\cite{RN217}, which differ in thermodynamic treatment. These advanced models are not directly comparable due to their different physical assumptions and numerical complexity. 
%add_revision
Moreover, agreement between implementations at experimentally relevant scales cannot be assumed, even if an agreement exists under a different set of conditions. For example, a companion NASA study~\cite{NASA_TM_2025} in which MOOSE was compared against PRISMS-PF under identical benchmark conditions showed significantly divergent results, despite both codes showing reasonable agreement on the idealized PFHub 3a.1 benchmark~\cite{wheeler2019pfhub}.
%add_revision
For benchmarking, it is critical to evaluate codes within a common framework. We therefore adopt the one-sided dilute alloy model of Echebarria et al.~\cite{RN32} as the shared reference,
%add_revision 
which incorporates an anti-trapping current~\cite{karma2001phase} to eliminate spurious solute trapping and recovers the sharp-interface limit via the matched asymptotic analysis~\cite{RN32,RN171,RN34}.
%add_revision
This model is implemented in PRISMS-PF and has been shown to accurately reproduce experimental observations of dendritic growth under microgravity~\cite{RN43,RN163}. Its assumptions of dilute alloy composition, negligible solid diffusivity, and local equilibrium are satisfied by the DECLIC-DSI-R benchmark system. 

We further design benchmark conditions that challenge numerical robustness by including interfacial instabilities in the initial conditions, features often avoided in conventional comparisons due to their nonlinear sensitivity. Typically, parameters and initial conditions are chosen to avoid interfacial instabilities, as they complicate comparisons between numerical implementations~\cite{RN168}. However, in this work, both codes implement the same underlying model, allowing us to focus on experimentally relevant phenomena in which instabilities play a central role. For instance, during directional solidification, an initially planar interface becomes unstable and breaks down under a small initial short-wavelength perturbation~\cite{RN59}, eventually evolving into dendritic or cellular structures. These structures then undergo competitive growth and selection dynamics, leading to the elimination of less favorable branches~\cite{RN89}. Benchmarking PF simulations during these processes is essential to the ICME framework, which relies on validated models to bridge experiments and materials design.

In this work, we present a quantitative benchmark comparing two state-of-the-art PF implementations: GPU-PF, a CUDA-based, finite-difference in-house research code optimized for GPUs and benchmarked with the DECLIC-DSI-R experiments; and PRISMS-PF, an open-source, MPI-parallelized finite element framework featuring adaptive mesh refinement and multilevel parallelism~\cite{RN194,RN210,RN211,RN212,RN214}. We focus on simulating three-dimensional, convection-free directional solidification of a binary alloy, a canonical testbed for PF modeling, where detailed microgravity experimental data from DECLIC-DSI-R program enables quantitative comparison. Both codes implement the same quantitative PF model for alloy solidification, incorporating anti-trapping currents and matched asymptotics to recover the sharp-interface limit under the thin-interface approximation. This study demonstrates the relative performance, scalability, and predictive agreement of both codes and provides a foundation for future integration of PF models with experimental workflows in alloy design under the ICME paradigm.

We benchmark the two selected codes in both two-dimensional (2D) and three-dimensional (3D) systems, namely 2D simulations of Aluminum-3wt\%Copper (Al-3 wt\% Cu) under high-velocity solidification conditions and large-scale 3D simulations of directional solidification in the organic alloy succinonitrile-0.46 wt\% camphor (SCN-0.46wt\%camphor) under microgravity. This opportunity is uniquely enabled by open-access flight data in the Physical Sciences Informatics Database~\cite{RN331}, where solidification occurs in the absence of convection, providing clean benchmark conditions unattainable on Earth. The GPU-PF code has already been validated against this dataset, offering fast and accurate predictions~\cite{RN43, song2023cell, mota2023influence,RN163}. By benchmarking PRISMS-PF under the same conditions, we assess whether an open-source tool can achieve similar fidelity and examine the associated computational cost. This comparison is critical as in-space manufacturing efforts expand and simulations play an increasingly central role in guiding experimental design. Establishing public tools capable of reproducing real spaceflight outcomes strengthens the broader ICME infrastructure for materials research in space.

\section{Model Formulations}
\label{sec:model}
The PF method is well-suited for tackling time-dependent free-boundary problems (TDFBPs), where the evolution of a material’s microstructure is governed by the motion of interfaces separating thermodynamically distinct phases (e.g., solid and liquid in solidification) or mechanically distinct states (e.g., broken and unbroken regions in fracture). A key feature of the PF method is the representation of an interface as a diffuse region of finite thickness $W_0$. PF variables, often referred to as order parameters, encode material states through their values. The phase fields are commonly defined over entire computational domains and are typically coupled to additional field variables. For example, a solute concentration field $c$~\cite{RN32} or a temperature field $T$~\cite{RN43} is required for modeling directional solidification. These fields evolve over time according to governing equations derived variationally from a free energy functional $\mathcal{F}$, which represents the total free energy of the system.

The benchmark problems presented in Section~\ref{sec:results} involve the solidification of binary alloys, where the PF variable $\phi$ captures the solid-liquid interface, and the solute concentration field $c$ tracks solute redistribution. The evolution of these fields is governed by a set of coupled partial differential equations (PDEs) that encode the thermodynamic driving forces and kinetic constraints of solidification. Material parameters, initial conditions, and boundary conditions are carefully prescribed to ensure physically realistic behaviors.

\subsection{Phase-Field Formulations}
\label{sub:formulation}
The time-dependent fields in this PF formulations are the phase field $\phi$ and the dimensionless supersaturation field $U$, which can be interpreted as a local chemical driving force for solidification. For the phase field, $\phi=-1$ denotes the liquid phase, $\phi=1$ denotes the solid phase. The contour of $\phi=0$ conventionally defines the solid-liquid interface. The supersaturation field $U$ is defined as
\begin{equation}
U=\frac{1}{1-k}\left[\frac{2c/c_l^0}{(1+k)-(1-k)h(\phi)}-1\right],
\end{equation}
where $k$ is the partition coefficient, and the function $h(\phi)$ is the phase-field interpolation function. In this study, we adopt the isothermal variational formulation (IVF) where $h(\phi) = \phi$~\cite{RN34}. Let $c_l$ and $c_s$ represent the concentration values in the liquid and solid phases, respectively. The parameter $c_l^0$ denotes the equilibrium concentration on the liquid side of the interface at the reference temperature $T_0$, while $c_s^0$ is the corresponding solid-side concentration. 

In this work, we take $T_0$ as the \textit{solidus} temperature, i.e., $T_0=T_{\textit{solidus}}$, so $c_s^0=c_\infty$, where $c_\infty$ is the uniform far-field concentration. We select an initial condition where the planar interface is placed at the \textit{liquidus} temperature ($T_{init}=T_{\textit{liquidus}}$), for which $c_l=c_\infty$. Local equilibrium ($c_s^0 = k\,c_l^0$) then gives $c_l^0 = c_\infty / k$. With this initial condition, the supersaturation field evaluates to $U = -1$ throughout the system. In addition, we apply the Frozen Temperature Approximation (FTA), which assumes a one-dimensional, fixed temperature gradient $G$, while the sample is pulled at a constant pulling velocity $V_p$. This approximation avoids explicitly solving the temperature field and is commonly used in PF simulations of directional solidification~\cite{RN43}. In the laboratory frame, the temperature field $T$ along the coordinate, $d$, that coincides with the direction of the temperature gradient is prescribed as
\begin{equation}
T = T_0 + G\left[d - V_p t\right], \quad \text{where } d = 
\begin{cases}
y, & \text{in 2D}, \\
z, & \text{in 3D}.
\end{cases}
\end{equation}
In the following, $z$ is used for simplicity.

As shown in Ref.~\cite{RN32}, the dimensionless evolution equation for the phase field, $\phi$, is
\begin{equation}
\begin{aligned}
f(k) a_s^2(\bm{\hat n})\,\partial_t \phi =& \nabla\cdot\left[a_s^2(\bm{\hat n}) \nabla \phi\right] \\ 
&+ \sum_{q = x,y,z} \partial_q \left[ |\nabla \phi|^2\, a_s(\bm{\hat n})\, \frac{\partial a_s(\bm{\hat n})}{\partial (\partial_q \phi)} \right] \\
&+ \phi - \phi^3 
- (1 - \phi^2)^2\, \lambda \left( \frac{\tilde z - \tilde V_p t}{\tilde l_T} + U \right),
\end{aligned}
\label{eq:phi}
\end{equation}
is obtained from the variational derivative of the free energy of the system. Here, $\tilde l_T=\Delta T/(GW_0)$ is the dimensionless thermal length, and $W_0$ is the the isotropic interface width. The term $\tilde z = z/W_0$ is the dimensionless position measured along the  direction of the temperature gradient, $G$, from a chosen reference point, and $\tilde V_p = V_p\tau_0/W_0$ is the dimensionless pulling velocity. The function $a_s(\bm{\hat n})$ encodes the anisotropy of the interface, $\lambda$ is the dimensionless coupling parameter, and $f(k)$ specifies how the relaxation time is chosen to match the sharp-interface limit (details given later).

The dimensionless supersaturation, $U$, evolves according to the continuity equation, $\partial_t U = D_L\nabla^2 U-\nabla\cdot\bm{j_{at}}$, where $D_L$ is the diffusion coefficient in the liquid, and $\bm{j_{at}}$ is the anti-trapping current
\begin{equation}
\begin{aligned}
\bm{j_{at}}=a(\phi)W(1-\phi)c_l^0e^u\frac{\partial \phi}{\partial t}\bm{\hat n}
, \quad a(\phi) = \frac{[h(\phi)-1][1-q(\phi)]}{\sqrt2 (\phi^2-1)}.
\end{aligned}
\end{equation}
Here, $h(\phi)=\phi$ and $q(\phi)=(1-\phi)/2$. The quantity, $u$, measures the departure of the chemical potential from its equilibrium value $\mu_E(T_0)$ at reference temperature $T_0$, and is related to the dimensionless supersaturation via $u=\ln\left[1+(1-k)U\right]$. In dimensionless form, the continuity equation becomes
\begin{equation}
\begin{aligned}
&[1+k-(1-k)\phi]\partial_t U = \\
&\nabla\cdot\left\{\tilde D(1-\phi)\nabla U+\frac{1}{\sqrt 2}[1+(1-k)U]\partial_t\phi\frac{\nabla\phi}{|\nabla\phi|}\right\}\\
&+[1+(1-k)U]\partial_t\phi,
\end{aligned}
\label{eq:U}
\end{equation}
where $\tilde D = D_L\tau_0/W_0^2$ is the dimensionless diffusion coefficient in the liquid, $\tau_0$ is the isotropic characteristic PF relaxation time.

For quasi-equilibrium solidification, the sharp-interface limit is recovered by choosing parameters such that the kinetic coefficient $\beta$ in the velocity-dependent Gibbs-Thomson condition vanishes~\cite{karma2001phase}. Following Eq.~(119) of Ref.~\cite{RN32}, $\beta$ is related to the phase-field parameters by
\begin{equation}
\beta=\frac{a_1\tau}{\lambda W}\left\{1-a_2\frac{\lambda W^2}{\tau}\frac{[1+(1-k)U]}{D_L}\right\}=0.
\end{equation}
%add_revision
Here, $\lambda=a_1W/d_0$ is the dimensionless coupling parameter, where $a_1, a_2$ are mapping constants obtained from the thin-interface asymptotic analysis. The term $d_0=\Gamma/[|m|(1-k)c_l^0]$ is the capillary length, where $\Gamma$ is the Gibbs-Thomson coefficient of the solid-liquid interface, $m$,  the liquidus slope, and $W$ and $\tau$ are the anisotropic interface width and relaxation time, respectively.
%add_revision
The mapping constants used in this study are $a_1\approx0.8839, a_2\approx0.6267$~\cite{RN34}.

Two approaches exist for selecting $f(k)$ and the relaxation time to ensure $\beta=0$~\cite{RN32}. The first approach is to use a temperature-dependent PF relaxation time
\begin{equation}
\tau=\tau_0f(k,z)=\tau_0[1-(1-k)(\tilde z-\tilde V_pt)/\tilde l_T].
\label{eq:zdeptau}
\end{equation}

The second approach is to use a $U$-dependent PF relaxation time
\begin{equation}
\tau=\tau_0f(k,U)=\tau_0[1+(1-k)U].
\label{eq:Udeptau}
\end{equation}
For simplicity and generality, the $U$-dependent form is adopted as the standard when benchmarking the two codes.

For anisotropy, we adopt the four-fold ``Kubic-Harmonic'' expansion with a single parameter $\varepsilon_4$~\cite{RN151} to parameterize the interfacial energy $\gamma(\bm{\hat n})=\gamma_0a_s(\bm{\hat n})$ and the interface width $W(\bm{\hat n})$. The normalized form is
\begin{equation}
a_s(\bm{\hat n})=\frac{W(\bm{\hat n})}{W_0} = \frac{\gamma(\bm{\hat n})}{\gamma_0}=(1 - 3\varepsilon_4) \left[ 1 + \frac{4\varepsilon_4}{1 - 3\varepsilon_4} \left( n_x^4 + n_y^4 + n_z^4 \right) \right].
\end{equation}
The above formulations correspond to the standard phase-field model without preconditioning.
In practice, when using large grid sizes, numerical stability can deteriorate. To address this, the GPU-PF employs a preconditioned phase field~\cite{RN215}, which improves the numerical stability under coarse spatial resolution. Details of this preconditioning method are provided in Section~\ref{sec:numerical}.

\subsection{Physical Parameters and System Overview}
\label{sub:paramters}

\begin{table}[t]
\centering
\renewcommand{\arraystretch}{1.3}
\resizebox{\Columnwidth}{!}{
    \begin{tabular}{clccr}
    \toprule
    \textbf{Symbol} & \textbf{Quantity} & \textbf{DSI-R~\cite{RN163}} & \textbf{Al-Cu} & \textbf{Unit} \\
    \midrule
    $c_0$ & Camphor concentration & 0.46 & 3.0 & $\mathrm{wt\%}$ \\
    $D_L$ & Diffusion constant in liquid & $2.7 \times 10^{-10}$ & $2.4 \times 10^{-9}$ & $\mathrm{m^2/s}$ \\
    $k$ & Partition coefficient & 0.1 & 0.14 & -- \\
    $m$ & Liquidus slope & -1.365 & -3.0 & $\mathrm{K/wt\%}$ \\
    $\Gamma$ & Gibbs--Thomson coefficient & $6.478 \times 10^{-8}$ & $2.4 \times 10^{-7}$ & $\mathrm{K\,m}$ \\
    $\epsilon_4$ & Crystalline anisotropy & 0.011 & 0.01 & -- \\
    $d_0$ & Capillary length & 0.0043 & 0.0115 & $\mathrm{\mu m}$ \\
    $G$ & Temperature gradient & 12.0 & 265.79 & $\mathrm{K/cm}$ \\
    $V_p$ & Pulling velocity & 6.0 & 18429.29 & $\mathrm{\mu m/s}$ \\
    $W_0$ & PF interface thickness & 1.265 & 0.098 & $\mathrm{\mu m}$ \\
    $\tau_0$ & PF time constant & 0.363 & $5.04 \times 10^{-5}$ & $\mathrm{s}$ \\
    \bottomrule
    \end{tabular}
}
\caption{Physical parameters used in the benchmark systems for SCN–camphor (DSI-R) and Al-Cu.}
\label{tab:parameters}
\end{table}
In this section, we present the model parameterization, outlining for both 2D and 3D cases. The computational domains, boundary conditions and initial conditions, including the initial planar front's location and perturbations, are described in detail in Section~\ref{sec:results}.

We consider two binary alloy systems for benchmarking: Al-3~wt\%~Cu under high-velocity solidification conditions and SCN-0.46~wt\% camphor under microgravity conditions from NASA’s DECLIC-DSI-R experiments. Their physical parameters are summarized in Table~\ref{tab:parameters}, including diffusion coefficients $D_L$, partition coefficients $k$, liquidus slopes $m$, and PF-specific quantities such as interface width $W_0$ and relaxation time $\tau_0$.

The Al-Cu benchmark was performed in a 2D setup with a quarter-circle solidification seed placed at one corner of the domain. This setup represents a high-speed solidification regime typical of high-velocity solidification processes, characterized by faster interface kinetics and steeper thermal gradients. This case was chosen as the publicly available example of the PRISMS-PF alloy solidification application to ensure it could be completed on a consumer-grade computer with limited cores in under 15 minutes. While outside the near-equilibrium regime for which the PF formulation is ideally suited, it serves as a practical baseline for verifying numerical agreement between codes and ensuring rapid benchmark convergence.

The DSI-R benchmark was initially conducted in a 2D configuration to enable rapid execution, with the initial interface position defined as a small amplitude perturbation around the liquidus temperature line. This setup was subsequently extended to three dimensions in two variants: one employing symmetry through no-flux (mirror) boundary conditions to reduce computational cost, and another using the full domain with no-flux boundaries but without symmetry reduction. The latter contains a sufficient number of dendrites (one full in the center and eight nearest neighbors) arranged in a simple cubic array to capture competitive growth, elimination, and primary spacing selection, representing the minimum domain size required in simulations to study the competitive growth of dendrites. In addition to enabling a more stringent performance assessment, comparing the two variants allows verification of the consistency of the symmetry-reduced approach.

% The DSI-R system presents a diffusion-limited regime ideal for benchmarking predictive PF models. Its low pulling velocity and thermal gradient produce quasi-steady-state dendritic arrays with primary spacing on the order of hundreds of microns, while eliminating convection through microgravity. In contrast, the Al-Cu benchmark reflects a high-speed solidification regime typical of AM processes, with faster interface kinetics and steeper thermal gradients. Together, these systems span a broad physical range that challenge both numerical stability and performance across different modeling frameworks.

\section{Numerical Method}
\label{sec:numerical}

In this section, we first outline the general numerical strategies applicable to phase-field models, followed by detailed descriptions of the specific implementations in PRISMS-PF and the finite-difference GPU code, GPU-PF. Section~\ref{sub:formulation} presented the nondimensionalized formulation suitable for numerical solution. With appropriate choices of the characteristic parameters $W_0$ and $\tau_0$, the update equations for the phase field, Eqs. \eqref{eq:phi}, and the supersaturation field, Eqs. \eqref{eq:U}, are scaled to be of order one, enabling efficient implementation using various numerical methods. In general, phase-field models can be discretized in space using the finite difference method (FDM), the finite element method (FEM), or the finite volume method (FVM). For time integration, explicit schemes are commonly used due to their computational efficiency. 

%delte_revision
% GPU-PF implements FDM on spatially uniform, fixed structured grids, combined with explicit Euler time stepping with a fixed time step. The simplicity of FDM minimizes computational overhead, making it highly efficient. In contrast, PRISMS-PF uses FEM on adaptive meshes, which are periodically remeshed based on a user-defined parameter. While it is more complex to implement and potentially more computationally demanding, FEM supports arbitrary geometries via unstructured meshes, enables higher-order accuracy, and integrates naturally with multiphysics coupling through modular extensions.  Moreover, PRISMS-PF employs the matrix-free finite element approach with sum-factorization and Gauss-Lobatto quadrature. This combination allows for efficient explicit time-stepping because it handles a trivially invertible diagonal “mass matrix”. 
% For both codes, convergence studies are conducted independently for the benchmarking cases described in Section~\ref{sub:paramters}, comparing the solid–liquid interface contours of the final solidified structures or steady-state dendrite arrays.
%delte_revision

\subsection{PRISMS-PF implementation}
\label{sec:prisms_pf}

PRISMS-PF is an open-source, high-performance framework for phase-field simulations of microstructure evolution~\cite{RN98}. 
%add_revision
It employs the FEM on adaptive meshes, which are periodically remeshed based on a user-defined parameter. While FEM is generally more complex and computationally demanding to implement than FDM, it supports arbitrary geometries via unstructured meshes, enables higher-order accuracy, and integrates naturally with multiphysics coupling through modular extensions. Moreover, PRISMS-PF employs the matrix-free finite element approach with sum-factorization and Gauss-Lobatto quadrature. This combination allows for efficient explicit time-stepping because it handles a trivially invertible diagonal “mass matrix”. 
%add_revision
PRISMS-PF is written in the C++ programming language and built upon the deal.II finite element library~\cite{RN224}. In addition to the features mentioned above, PRISMS-PF has three levels of parallelization: distributed memory, task-based threading, and vectorization, which makes it highly scalable. Moreover, the use of quadratic or cubic elements is advantageous for problems that require higher resolution at the interface. 

For the implementation of the model~\cite{RN311}, we must obtain the time-discretized version of Eqs.~\eqref{eq:phi} and~\eqref{eq:U}, with $\tau$ given by Eq. \eqref{eq:Udeptau}, and then obtain their weak form. For convenience, we introduce an auxiliary field $\xi(\phi,U)$ as
\begin{equation}
\begin{aligned}
\xi = & \nabla\cdot\left[a_s^2(\bm{\hat n}) \nabla \phi\right] \\ 
&+ \sum_{q = x,y,z} \partial_q \left[ |\nabla \phi|^2\, a_s(\bm{\hat n})\, \frac{\partial a_s(\bm{\hat n})}{\partial (\partial_q \phi)} \right] \\
&+ \phi - \phi^3 
- (1 - \phi^2)^2\, \lambda \left( \frac{\tilde z - \tilde V_p t}{\tilde l_T} + U \right).
\end{aligned}
\label{eq:xi}
\end{equation}
We then rewrite Eqs.~\eqref{eq:phi} and~\eqref{eq:U} as
\begin{equation}
\partial_t\phi = \frac{\xi}{\tau a_s^2(\bm{\hat n})},
\label{eq:phi_reform}
\end{equation}
and 
\begin{equation}
\begin{aligned}
\partial_t U =& 
\left\{\nabla\cdot\left[\tilde D(1-\phi)\nabla U+\frac{1}{\sqrt 2}[1+(1-k)U]\frac{\xi}{\tau a_s^2(\bm{\hat n})}\frac{\nabla\phi}{|\nabla\phi|}\right] \right.\\
&\left.+[1+(1-k)U]\frac{\xi}{\tau a_s^2(\mathbf{\hat{n}})}\right\}[1+k-(1-k)\phi]^{-1}.
\end{aligned}
\label{eq:U_reform}
\end{equation}

The weak formulations of Eqs.~\eqref{eq:xi}-\eqref{eq:U_reform} have the following general form
\begin{equation}
\int_\Omega\omega\Psi^{i+1}dV = \int_\Omega\omega s_\Psi^i dV + \int_\Omega\nabla\omega\cdot\bm{\mathrm{v}}_\Psi^{\,i} dV,
\label{eq:general_weak}
\end{equation}
where $\Psi$ represents any of the fields, $\phi$, $U$, or $\xi$. The integral is over the domain volume, $\Omega$, and  $\omega$ is an arbitrary test function that vanishes on the domain boundaries, i.e., $\omega=0$ along $\partial \Omega$. The superscript in Eq.~\eqref{eq:general_weak} denotes the time increment ($i$ or $i+1$). All terms on the right-hand side (RHS) depend on fields evaluated at time $t^i$. The RHS terms $s_\Psi^i$ and $\bm{\mathrm{v}}_\Psi^{\,i}$ are scalar and vector expressions that must be explicitly codified in PRISMS-PF and passed to the solver. They are given by
\begin{equation}
\begin{aligned}
s_\xi^i = \phi^i - (\phi^i)^3 
- \left[1 - (\phi^i)^2\right]^2\, \lambda \left( \frac{\tilde z - \tilde V_p t^i}{\tilde l_T} + U^i \right), 
\end{aligned}
\end{equation}
\begin{equation}
\begin{aligned}
\bm{\mathrm{v}}_\xi^{\,i} = -\bm{A}(\nabla\phi^i),
\end{aligned}
\end{equation}
\begin{equation}
\begin{aligned}
s_\phi^i = \phi^i + \frac{\xi^i}{\tau^i a_s^2(\bm{\hat n}^{\,i})}\Delta t, 
\end{aligned}
\end{equation}
\begin{equation}
\begin{aligned}
\bm{\mathrm{v}}_\phi^{\,i} = 0,
\end{aligned}
\end{equation}
\begin{equation}
\begin{aligned}
s_U^i = & U^i + \Delta t \left[\frac{1+(1-k)U^i}{1+k-(1- k)\phi^i}\right]\frac{\xi^i}{\tau^i a_s^2(\bm{\hat n}^{\,i})}\\
& - \frac{\Delta t}{\left[1+k-(1-k)\phi^i\right]^2}\left\{\tilde{D}(1-\phi^i)\nabla U^i\cdot \nabla \phi^i \right.\\
&\left. +\frac{1}{\sqrt{2}}\left[1+(1-k)U^i\right]\frac{\xi^i}{\tau^i a_s^2(\bm{\hat n}^{\,i})}\left|\nabla \phi^i\right|\right\},
\end{aligned}
\end{equation}
and
\begin{equation}
\begin{aligned}
\bm{\mathrm{v}}_U^{\,i} = & -\frac{\Delta t}{1+k-(1- k)\phi^i}\left\{ \tilde{D}(1-\phi^i)\nabla U^i \right.\\
&\left. +\frac{1}{\sqrt{2}}\left[1+(1-k)U^i\right]\frac{\xi^i}{\tau^i a_s^2(\bm{\hat n}^{\,i})}\frac{\nabla \phi^i}{\left|\nabla \phi^i\right|}\right\},
\end{aligned}
\end{equation}
where the components of vector $\bm{A}(\nabla\phi^i)$ are given by
\begin{equation}
A_q = a^2_s(\bm{\hat n}^{\,i})\partial_q\phi^i+|\nabla \phi^i|^2\, a_s(\bm{\hat n}^{\,i})\, \frac{\partial a_s(\bm{\hat n}^{\,i})}{\partial (\partial_q \phi^i)},
\end{equation}
and $\Delta t$ is the time increment.

Employing the first-order explicit Euler time-integration scheme described above, Eqs.~\eqref{eq:xi}-\eqref{eq:U_reform} are solved in PRISMS-PF with an adaptive mesh consisting of second-order Lagrange elements and Gauss-Lobatto quadrature. All simulations were run with PRISMS-PF v2.4 and deal.II 9.5.2.

\subsection{GPU-PF implementation}
\label{sub:gpu_pf}
GPU-PF is a finite-difference solver designed for structured grids, implemented in CUDA~C++ for execution on both single- and multi-GPU architectures. 
%add_revision
GPU-PF implements FDM on spatially uniform, fixed structured grids, combined with explicit Euler time stepping with a fixed time step. The simplicity of FDM minimizes computational overhead, making it highly efficient.
%add_revision
Spatial derivatives are evaluated using second-order central differences, while time integration employs a first-order explicit Euler scheme. For multi-GPU execution, domain decomposition is used in combination with CUDA streams to asynchronously launch kernel updates and perform halo exchanges between subdomains. All kernel operations on each GPU, including field updates and boundary conditions, are overlapped using non-blocking streams to maximize concurrency and throughput.

% preconditioned PF
To extend numerical stability at larger grid spacings, we replace the standard phase field $\phi$ with a nonlinear transformation to a preconditioned phase field $\psi$~\cite{RN215}
\begin{equation}
\phi=\tanh\left(\frac{\psi}{\sqrt2}\right).
\end{equation}
The evolution equations for the preconditioned phase field $\psi$ and dimensionless supersaturation $U$ can be written as~\cite{RN38,RN45}:
\begin{equation}
\begin{aligned}
\tilde{F}_1 a_s^2(\bm{\hat n}) \frac{\partial \psi}{\partial t} &= a_s^2(\bm{\hat n}) \left( \nabla^2 \psi - \phi \sqrt{2}|\nabla \psi|^2 \right) + \nabla[a_s^2(\bm{\hat n})] \cdot \nabla \psi \\
&\quad + \sum_q \partial_q \left( |\nabla \psi|^2 a_s(\bm{\hat n}) \frac{\partial a_s(\bm{\hat n})}{\partial (\partial_q \psi)} \right) \\
&\quad + \sqrt{2} \left[\phi - \lambda (1 - \phi^2)\tilde{F}_2 \right],
\end{aligned}
\end{equation}
and 
\begin{equation}
\begin{aligned}
\left[1 + k - (1 - k)\phi\right] \partial_t U &= \tilde{D} \nabla \cdot \left( [1 - \phi] \nabla U \right) \\
&+ \nabla \cdot \left\{ [1 + (1 - k)U] \frac{(1 - \phi^2)}{2} \frac{\partial \psi}{\partial t} \frac{\nabla \psi}{|\nabla \psi|} \right\} \\
&+ [1 + (1 - k)U] \frac{(1 - \phi^2)}{\sqrt{2}} \frac{\partial \psi}{\partial t},
\end{aligned}
\end{equation}
with definitions of two thermal functions
\begin{equation}
\tilde{F}_1(z,t) = \left[ 1 - (1 - k) \frac{\tilde{z} - \tilde{V}_p t}{\tilde{l}_T} \right],
\end{equation}
and 
\begin{equation}
\tilde{F}_2(z,t) = \left[ U + \frac{\tilde{z} - \tilde{V}_p t}{\tilde{l}_T} \right].
\end{equation}
Here, $\tilde F_1(z,t)$ corresponds to the choice of relaxation time that ensures asymptotic matching to the sharp-interface limit, as given in Eqs.~\eqref{eq:zdeptau} and \eqref{eq:Udeptau}. 
In GPU-PF, Eq.~\eqref{eq:zdeptau} was adopted~\cite{RN38,RN45}, whereas PRISMS-PF employs the U-dependent form, as in Eq.~\eqref{eq:Udeptau}. Both forms are physically equivalent. For benchmarking with PRISMS-PF and to minimize numerical differences between implementations, the thermal function $\tilde{F}_1(z,t)$ was adjusted in this study to
\begin{equation}
\tilde{F}_1(z,t) = \left[ 1 + (1 - k) U \right]
\end{equation}

% isotropic discretization in 2D
For the 2D benchmark, we implemented isotropic finite-difference discretization~\cite{RN38} on structured Cartesian grids to mitigate spurious lattice anisotropy effects. The GPU-PF code discretizes key differential operators, i.e., the Laplacian, solutal flux divergence, and anti-trapping current, with rotational invariance at order $\mathcal{O}(\Delta x^2)$, achieved by combining finite-difference stencils along $\langle 10 \rangle$ and $\langle 11 \rangle$ directions using a $2{:}1$ weight ratio. The preconditioned phase field $\psi$ introduces additional nonlinear gradient terms, such as $|\nabla \psi|^2$, which are also treated with isotropic discretization to preserve accuracy and stability. 

Notably, the anti-trapping current, originally in the form $\nabla \cdot (\alpha \nabla \beta / |\nabla \beta|)$, is approximated by assuming a stationary $\tanh$ profile. This transforms it into the divergence of a flux $\nabla \cdot (\alpha \nabla \beta)$, making it compatible with the same isotropic discretization framework. This isotropic formulation significantly improves the accuracy of dendritic tip dynamics, especially in systems where the dendrite growth direction is misaligned with the computational grid, such as the Al-Cu benchmark case, where a side branch grows close to the $\langle 11 \rangle$ direction, which is more prone to grid-induced distortion with anisotropic stencils. However, in three dimensions, the isotropic stencil becomes prohibitively complex due to the need for extended neighborhoods (e.g., in the〈110〉 and 〈111〉 directions). Therefore, in 3D GPU-PF simulations, we retain the standard anisotropic discretization to preserve computational efficiency and simplicity.

\section{Results and Discussion}
\label{sec:results}

This section presents a side-by-side comparison of GPU-PF and PRISMS-PF for directional solidification in benchmark systems representative of high-velocity solidification and microgravity experiments. The benchmarks include 2D Al-Cu and SCN-camphor systems, as well as 3D SCN-camphor systems, selected to test both physical fidelity and numerical performance. The analysis examines agreement in predicted morphologies and growth characteristics, along with convergence behavior and computational efficiency. Results are organized by benchmark setups to isolate the effects of numerical formulation from physical modeling assumptions, with attention to cases that challenge numerical robustness and large-scale simulations relevant to spaceflight experiments.

\subsection{2D Simulation: Al-3wt\%Cu under High-velocity Solidification Conditions}
\label{sec:2DAlcu}

\begin{figure}[t]%% placement specifier
\centering%% For centre alignment of image.
\includegraphics[width=0.5\textwidth]{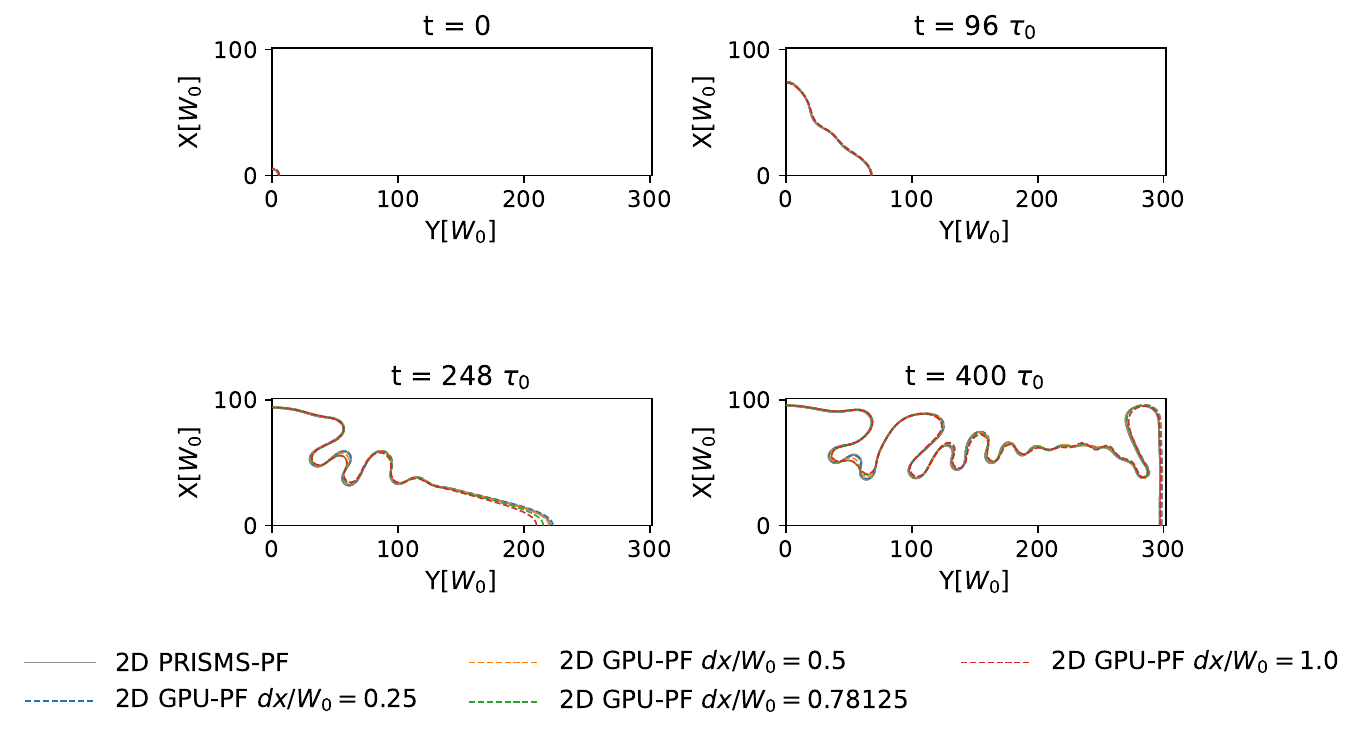}
%% Use \caption command for figure caption and label.
\caption{Comparison of $\phi = 0$ solid-liquid interface contours at four different time slices from two simulations: the PRISMS-PF finite element implementation (grey) and the GPU-PF finite difference code (colored dash lines). The excellent agreement confirms consistent interface evolution between the two codes at this intermediate stage.}\label{fig1}
\end{figure}

The Al-Cu benchmark case is based on PRISMS-PF’s default example configuration. It represents a quarter-circle solidified seed in a $29.5~\mu\mathrm{m} \times 14.7~\mu\mathrm{m}$ rectangular domain with no-flux boundaries. The interface is initialized with an undercooling offset of $U_{\mathrm{off}}=0.9$ near the liquidus temperature. For reference, $U_{\mathrm{off}}=0$ corresponds to the initial interface placed at the solidus temperature and $U_{\mathrm{off}}=1.0$ to the liquidus temperature. The simulation time was set to $400~\tau_0\approx 0.02 \mathrm{s}$. The dimensionless parameters are as follows: $\lambda=20.0$, $\tilde D=12.53$, $\tilde V_p=9.45$, $\tilde l_T=21175.72$, $\Delta x/W_0=0.78125$, $\Delta t/\tau_0=0.002$. This choice of $\Delta t$, as well as all other $\Delta t$ choices, satisfies the Courant-Friedrichs-Lewy (CFL) condition,  $\Delta t \le (\Delta x)^2/(2d\tilde D)$, where $d$ is the simulation dimensionality. Figure~\ref{fig1} shows a comparison of the solid-liquid interface evolution between the reference PRISMS-PF simulation and the GPU-PF simulation at four representative time slices, from a small circular crystal seed to an almost completely solidified sample. The $\phi = 0$ contours conventionally define the solidification fronts, capturing the morphological evolution of the quarter-seed dendrite. The initial condition shows a smooth quarter-circle shape, while subsequent frames illustrate tip-splitting, side-branching, and overall anisotropic growth. 

With second-order elements, at $\Delta x/W_0=0.78125, \Delta t/\tau_0=0.002$, the PRISMS-PF simulation was found to be nearly mesh-converged. The GPU-PF code, with isotropic discretization, exhibits excellent agreement with PRISMS-PF at all stages at around $\Delta x/W_0=0.25, \Delta t/\tau_0=0.002$, including fine-scale morphological features, validating the GPU implementation against the converged finite element result. In this case, the good agreement also reflects the advantages in using the isotropic discretization in GPU-PF, which mitigates lattice anisotropy effects of the standard finite-difference discretization.

\subsection{2D Simulation: SCN-0.46wt\% Camphor under Microgravity (DSI-R)}
\label{sec:2DDSIR}
The comparison is extended to parameters of the DECLIC-DSI-R flight experiments (Table~\ref{tab:parameters}). This case corresponds to a transparent succinonitrile-0.46 wt\% camphor alloy directionally solidified at a low isothermal velocity ($V_p=6~\mathrm{\mu m/s}$) and a small thermal gradient ($G=12~\mathrm{K/cm}$). The simulation domain spans $3827~\mu\mathrm{m} \times 637.8~\mu\mathrm{m}$. A sinusoidal perturbation initiates interface instability
\begin{equation}
% y_{\mathrm{int}} = y_{\mathrm{int},0} + 0.5\mathcal{A}\left[1 + \cos\left(k_q \pi(L_x - x)/L_x\right)\right]
y_{\mathrm{int}} = y_{\mathrm{int},0} + 0.5\mathcal{A}\left[1 + (-1)^n \cos\left(\kappa x\right)\right]
\end{equation}
where $y_{\mathrm{int},0} = 379.28~W_0$, $\mathcal{A} = 10~W_0$, $\kappa =n\pi /L_x$ and $n=3$. Unlike the Al-Cu case, the initial interface here is placed directly at the liquidus temperature without a fixed undercooling offset, so the solidification velocity starts from zero and increases as the interface recoils and subsequently destabilizes. 

\begin{figure}[t]%% placement specifier
\centering%% For centre alignment of image.
\includegraphics[width=0.35\textwidth]{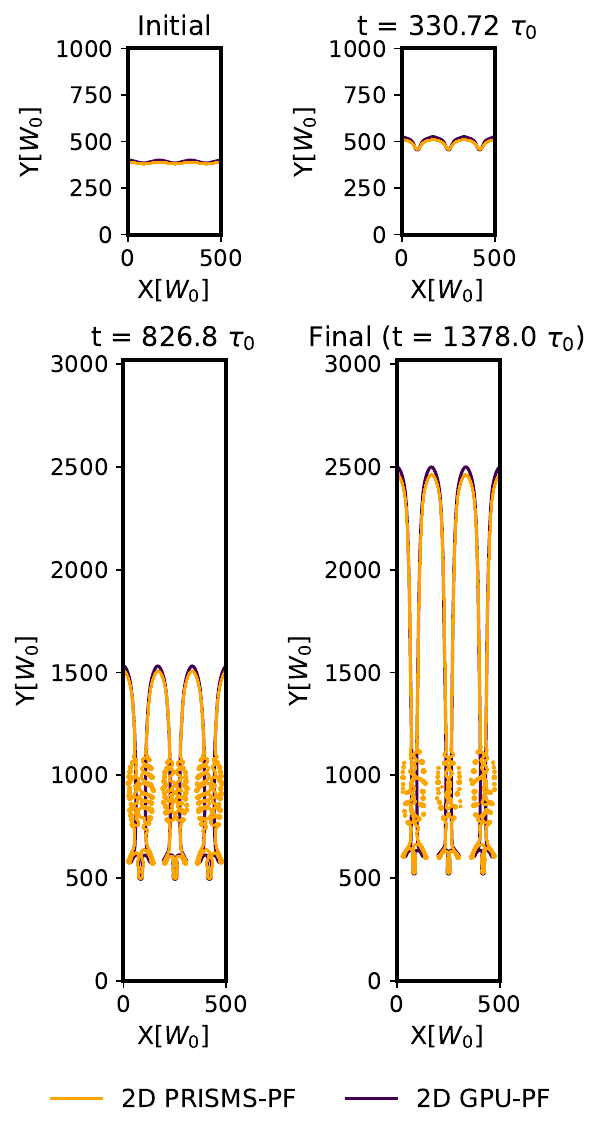}
%% Use \caption command for figure caption and label.
\caption{Comparison of $\phi = 0$ solid-liquid interface contours at four different time slices from two simulations: the PRISMS-PF finite element implementation (orange) and the GPU-PF finite difference code (purple). The excellent agreement confirms consistent interface evolution between the two codes at this intermediate stage.}\label{fig2}
\end{figure}

The domain length is chosen large enough to capture the later stages of interface evolution in the absence of the treadmill functionality~\cite{RN329,RN330}. The simulation time is set to $1378.0~\tau_0 \approx 500.0~\mathrm{s}$, during which the interface reaches $y\approx 2500~\mathrm{\mu m}$. This leaves more than $5 l_D$ ($l_D = D_L / V_p = 45~\mathrm{\mu m}$) of liquid ahead of the front (see Fig. \ref{fig2}). The remaining domain ahead ensures accurate solute diffusion and prevents artificial boundary effects. The dimensionless parameters are $\lambda=97.58$, $\tilde D=61.15$, $\tilde V_p=1.718$, $\tilde l_T=3725.73$, with spatial and temporal resolutions of $\Delta x/W_0=0.984375$ and $\Delta t/\tau_0=5\times10^{-4}$.

Figure~\ref{fig2} overlays the solid-liquid interface contours from GPU-PF and PRISMS-PF simulations at four representative times, illustrating the temporal evolution of a sinusoidally perturbed planar front. The contours capture the onset of planar interface breakdown, cellular instabilities, and the formation of a cellular array. Both implementations closely track the interface progression, accurately resolving tip splitting and cellular growth with high consistency. The overlap is nearly perfect, except deep in the liquid groove region, where a transient side-branching instability occurs during the initial destabilization phase. This deviation has been shown to have a negligible influence on the subsequent morphological evolution, owing to the fact that the growth kinetics of the dendrite tip dominate at later stages~\cite{RN53}. The near-perfect agreement between the GPU-PF (purple) and PRISMS-PF (orange) interfaces confirms numerical convergence.

\subsection{3D Simulation: SCN-Camphor under DSI-R Conditions}
\label{sec:3DDSIR}
The above 2D simulation comparison was extended to three dimensions using the same parameters as in Table~\ref{tab:parameters}; the 3D simulations thus more accurately reflect the conditions in the microgravity experiments. This benchmark evaluates both the convergence and the performance, providing a basis to estimate the computational resources needed to model flight experiments on a full sample scale using publicly available PF codes. 

The 3D domain is $194 \times 194 \times 2332~\mu\mathrm{m}^3$. Two resolutions were tested: $\Delta x/W_0 = 0.8$ and $1.2$, mapped to $128\times128\times1536$ and $192\times192\times2304$ grid points with corresponding time step sizes of $\Delta t= 2 \times 10^{-4} ~\tau_0$ and $\Delta t= 5 \times 10^{-4} ~\tau_0$, respectively, in GPU-PF. In PRISMS-PF, the minimum mesh spacing was selected to provide comparable element size. While GPU-PF's preconditioned formulation is stable up to $\Delta t= 3.5 \times 10^{-3} ~\tau_0$ for $\Delta x/W_0=1.2$, approaching but below the CFL limit, $\Delta t \le (\Delta x)^2 / (6 \tilde{D})= 3.9 \times 10^{-3}~\tau_0$, we did not use this value to allow comparison with PRISMS-PF.

\begin{figure}[t]%% placement specifier
\centering%% For centre alignment of image.
\includegraphics[width=0.5\textwidth]{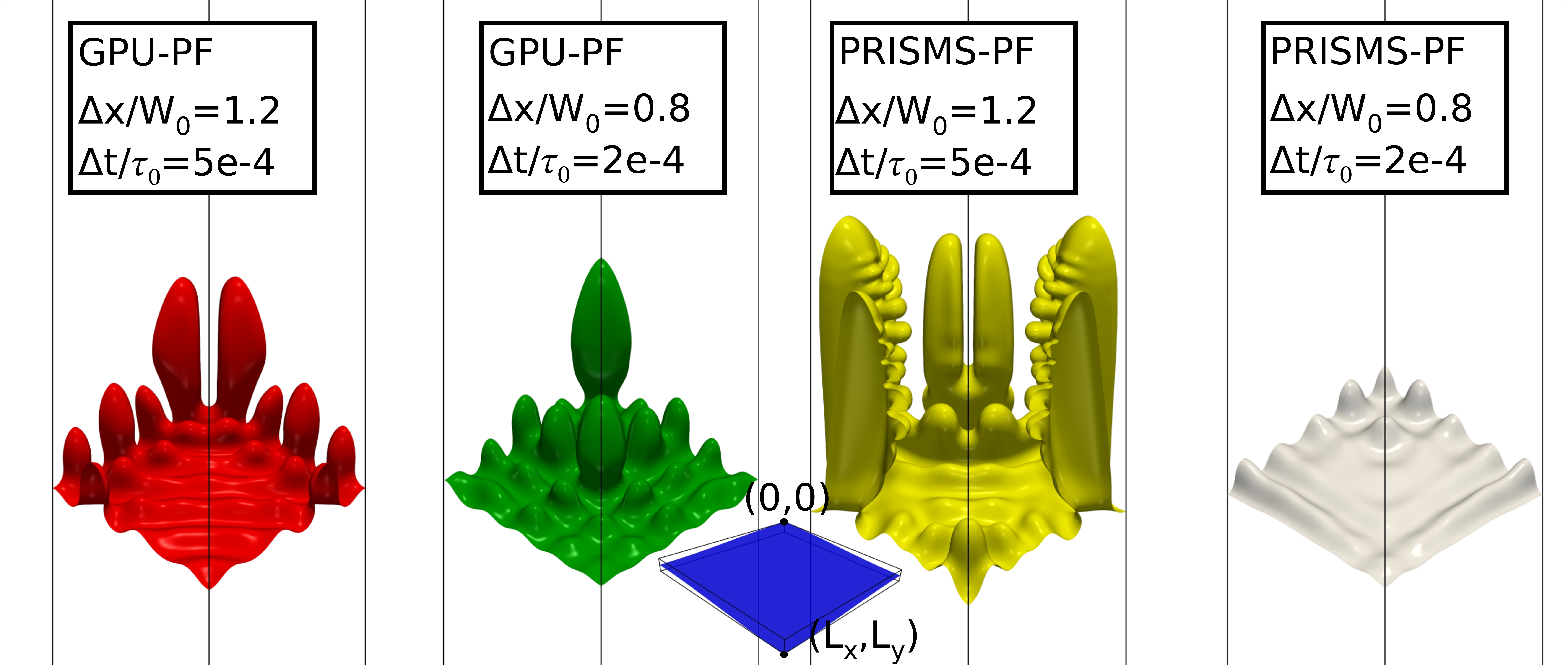}
%% Use \caption command for figure caption and label.
\caption{$\phi = 0$ solid-liquid interface contours at $t = 400\tau_0$ with long-wavelength perturbations ($n = 1$) for four cases: two spatial resolutions and the corresponding time steps sizes each for GPU-PF (left two) and PRISMS-PF (right two). Middle: the initial position of the perturbed interface.}\label{fig3}
\end{figure}

The general form of the initial position of the interface (also used for the remaining 3D benchmark simulations) can be expressed as
\begin{equation}
\begin{aligned}
% z_{\mathrm{int}}& = z_{\mathrm{int},0}\\
% &+\frac{1}{4}\mathcal{A} \left[ 2 + \cos\left(k_q\pi\left(1 - \frac{x}{L_x}\right)\right) + \cos\left(k_q\pi\left(1 - \frac{y}{L_y}\right)\right) \right]\\
% &+\frac{1}{8}\mathcal{B}\left[ \cos\left(\frac{2\pi x}{L_x}\right) + \cos\left(\frac{2\pi y}{L_y}\right) \right].\\
z_{\mathrm{int}}& = z_{\mathrm{int},0}+{\mathcal{A}_1}\left[ 2 + (-1)^{n}\cos\left(\kappa_1 x\right) + (-1)^{n}\cos\left(\kappa_1 y\right)  \right]\\
&+{\mathcal{A}_2}\left[ \cos\left(\kappa_2 x\right) + \cos\left(\kappa_2 y\right) \right],\\
\end{aligned}
\label{eq:perturbation}
\end{equation}
where $z_{\mathrm{int},0}$ is the initial position of the unperturbed planar interface, $n$ is an integer, $\kappa_1=n\pi/L_x=n\pi/L_y$ and $\kappa_2=2\pi/L_x=2\pi/L_y$ ($L_x=L_y$) are the wave numbers of two perturbations of different wavelengths ${2\pi}/{\kappa_1}$, ${2\pi}/{\kappa_2}$ with amplitudes $\mathcal{A}_1$, $\mathcal{A}_2$, respectively. 

% \subsubsection{Case 1: Chaotic Regime}
We first tested the two codes with a single long wavelength perturbation ($\mathcal {A}_1=6.09525~W_0$, $\mathcal {A}_2=0$, $n=1$ and $\kappa_1=\pi/L_x$, wavelength $2L_x$). Figure~\ref{fig3} shows the interface morphologies at $t=400~\tau_0$ for the two sets of resolution parameters, (1) $\Delta x/W_0=0.8, \Delta t/\tau_0=2\times 10^{-4}$ and (2) $\Delta x/W_0=1.2, \Delta t/\tau_0=5\times 10^{-4}$. The differences between the results with two resolutions are significant for both the PRISMS-PF and GPU-PF codes. Furthermore, the results from GPU-PF and PRISMS-PF are also vastly different, even for parameter set (1), which has a higher resolution finer than the resolution previously shown to yield converged results~\cite{RN163}. These differences can be explained as follows. The long wavelength modulation ($n=1$) is unstable and excites shorter wavelength modes not initially present. These modes grow into small bumps in the interface that trigger the competitive growth of cell/dendrite tips and secondary branches. Nonlinear interactions between these modes of different wavelengths lead to a complex chaotic evolution. Thus, the simulated interface morphology is strongly dependent on the initial conditions defined by Eq.~\eqref{eq:perturbation}, with small numerical differences being exponentially amplified by positive Lyapunov exponents. The dynamics governing these simulations are analogous to those arising from the three coupled ordinary differential equations that describe the chaotic Lorenz attractor. Despite the fact that both systems have deterministic equations, their time-evolution is exponentially diverging due to sensitivity to the initial conditions, rendering the comparison between the different simulations ill-posed.

\begin{figure*}[t]%% placement specifier
\centering%% For centre alignment of image.
\includegraphics[width=0.99\textwidth]{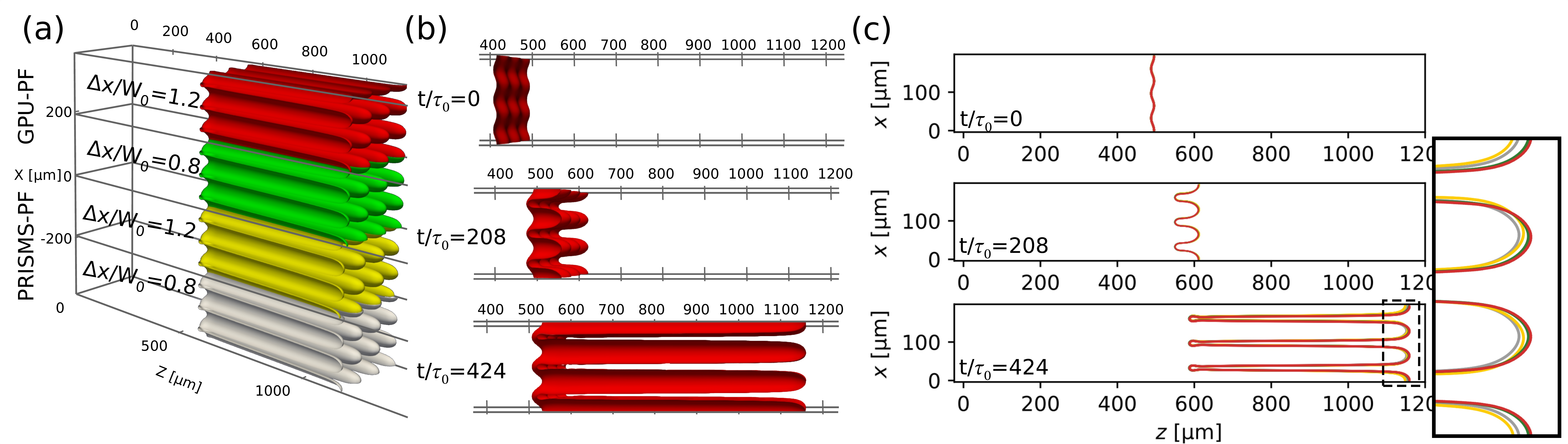}
%% Use \caption command for figure caption and label.
\caption{Comparison of 3D solid-liquid interface evolution between GPU-PF and PRISMS-PF under DECLIC-DSI-R benchmark conditions. The domain size is $(L_x,\,L_y,\,L_z)=(0.194~\mathrm{mm},\,0.194~\mathrm{mm},\,2.332~\mathrm{mm})$. 
(a) 3D interface contours at $t/\tau_0 = 424$. The dendrite morphologies show strong agreement across codes and resolutions. Red: GPU-PF simulation with $\Delta x/W_0 = 1.2$ and $\Delta t/\tau_0 = 5\times10^{-4}$; Green: GPU-PF simulation with $\Delta x/W_0 = 0.8$ and $\Delta t/\tau_0 = 5\times10^{-4}$; Yellow: PRISMS-PF simulation with $\Delta x/W_0 = 1.2$ and $\Delta t/\tau_0 = 5\times10^{-4}$; Grey: PRISMS-PF simulation with $\Delta x/W_0 = 0.8$ and $\Delta t/\tau_0 = 2\times10^{-4}$.
(b) Temporal evolution of the GPU-PF interface with $\Delta x/W_0=1.2$, showing $\phi=0$ contours at $t/\tau_0 = 0$, $208$, and $424$. A sinusoidally perturbed planar front grows into uniformly spaced dendrites through cellular instability.
(c) Cross-sectional views of the $x$–$z$ interface profile at $y=L_y/3$ for all simulations at $t/\tau_0 = 0$, $208$, and $424$. The right panel shows a zoomed-in view of the final front position at $t/\tau_0 = 424$, confirming excellent quantitative agreement across codes and resolutions. The color scheme follows that of (a).
}
\label{fig4}
\end{figure*}

% \subsubsection{Case 2: Pure Short-Wavelength Perturbation}
To avoid the chaotic nonlinear behavior of the example presented above, we varied the number, $n$, of the initial condition to identify cases where the interface dynamics are non-chaotic on the time scale of the simulation. We examine the evolution with the parameters $\mathcal {A}_1=12.1905~W_0$, $\mathcal{A}_2=0$ and $n=6$, which corresponds to a larger wavenumber, $\kappa_1=6\pi/L_x$. The choice $n=6$ corresponds to a shorter initial wavelength of $L_x/3\approx 66~\mathrm{\mu m}$, slightly above the solute diffusion length $l_D=D_L/V_p\approx 45~\mathrm{\mu m}$. The same set of space and time resolution values from the example above were used for the PRISMS-PF and GPU-PF simulations. As shown in Figure~\ref{fig4}, the results at $t/\tau_0=208$ and $424$ exhibit negligible dependence on grid resolution, thereby justifying a meaningful comparison between the codes. Excellent agreement is observed between the PRISMS-PF and GPU-PF results, with both simulations yielding nearly identical periodic dendrite arrays. One caveat associated with this set of initial condition parameters is that, although not observed in the simulations, the dendritic array is in an unstable state because dendrite primary spacing, $\Lambda$, falls outside the reported stable range. This range is known to be limited by cell elimination when $\Lambda$ is smaller than a minimum value, $\Lambda_{\text{min}}$, and by tertiary branching when $\Lambda$ exceeds a maximum value, $\Lambda_{\text{max}}$~\cite{RN89}. The previously reported stability band of the same system extends from $\Lambda_{\text{min}}=133~\mathrm{\mu m}$ to $\Lambda_{\text{max}}=316~\mathrm{\mu m}$~\cite{RN163}. The primary spacing obtained from the simulations for this benchmark case is $\Lambda\approx65~\mathrm{\mu m}$, which is smaller than the reported minimum spacing. In the simulations, the dendrites remain stable because the symmetry of the initial condition yields a perfectly periodic array that would need to be perturbed to trigger cell elimination, which is difficult given the domain size that is comparable to the stable primary spacing. Both GPU-PF and PRISMS-PF successfully sustain this solution throughout the simulation time, confirming the numerical consistency of the codes. However, in any real experimental system, thermal fluctuations, fluid flow, and other ambient perturbations would drive the interface away from this unstable equilibrium and toward a configuration with larger primary spacing. 
  
\begin{figure*}[t]%% placement specifier
\centering%% For centre alignment of image.
\includegraphics[width=1.0\textwidth]{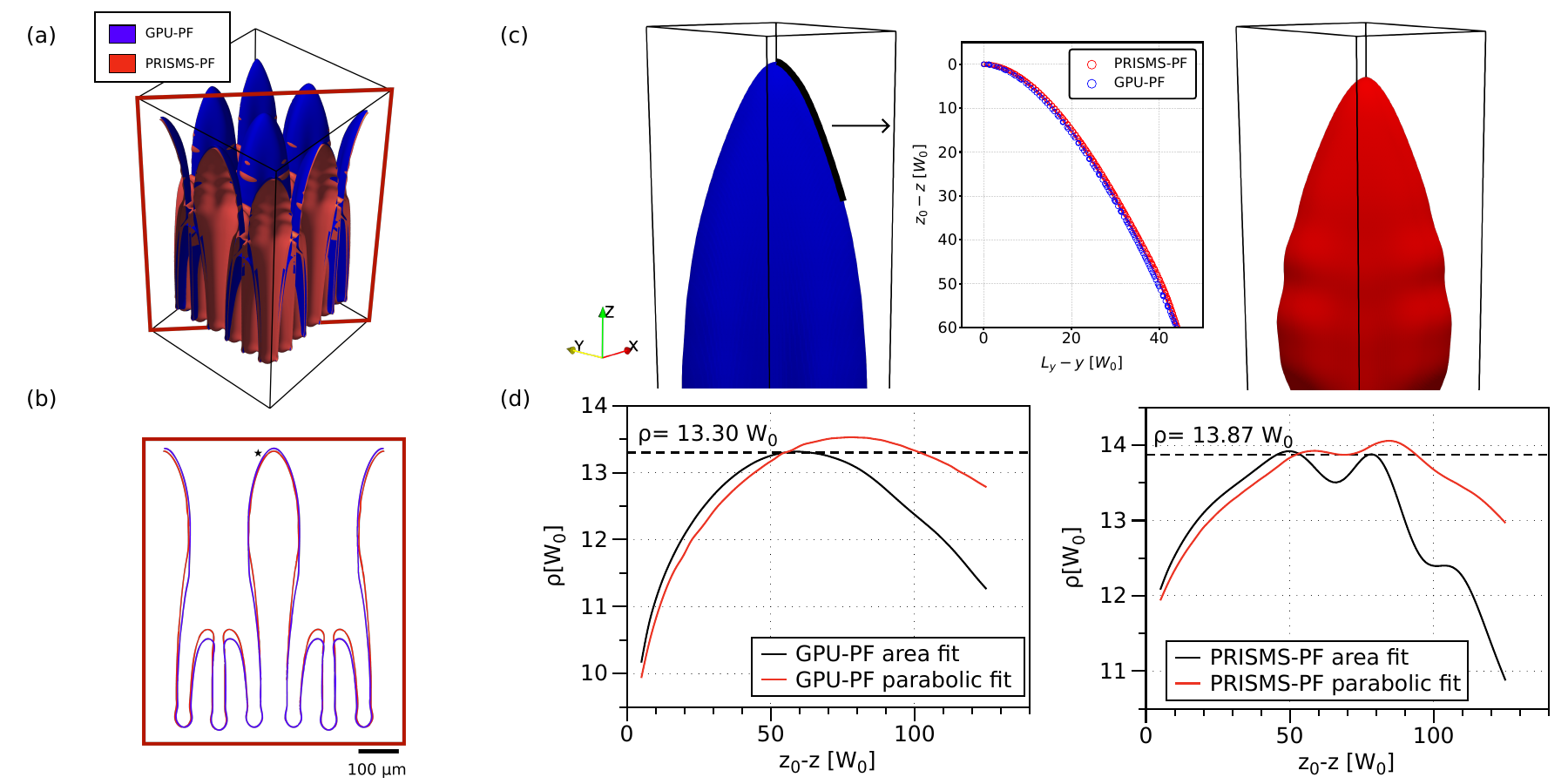}
%% Use \caption command for figure caption and label.
\caption{Interface contour and cross-section for both GPU-PF and PRISMS-PF simulations at $t=424~\tau_0$ using the composite initial condition defined in Eq.~\eqref{eq:perturbation}, with $\mathcal{A}_1=12.1905~W_0$, $\mathcal{A}_2=12.1905~W_0$ and $n=6$. The resolution $\Delta x/W_0=1.2$ and $\Delta t/\tau_0=5\times 10^{-4}$ was used for GPU-PF, and a comparable element size was used in PRISMS-PF. 
(a) 3D interface contours showing strong morphological agreement between the two codes. 
(b) Cross-sectional view of the $y-z$ interface profile at $x=L_x/2$, confirming near-identical dendrite tip positions and spacings. The central dendrite marked with $\star$ in panel (b) is used for the quantitative tip radius analysis shown in panels (c) and (d).
(c) Longitudinal 2D cross section profile near the tip of the central dendrite for both codes, with GPU-PF (blue) and PRISMS-PF (red) showing excellent overlap near the tip. 
(d) Local tip radius $\rho$ as a function of distance from the tip $z_0-z$, extracted using area fit (black) and parabolic fit (red) methods for GPU-PF (left) and PRISMS-PF (right). The tip radius is determined at the crossover point of the two fitting methods, yielding $\rho=13.30~W_0$ for GPU-PF and $\rho=13.87~W_0$ for PRISMS-PF, a difference of $4.3\%$}
\label{fig5}
\end{figure*}

% \subsubsection{Case 3: Composite Perturbation}
To obtain a result that yields dendrites with a stable primary spacing, we perform an additional comparison of the two codes using an initial condition that leads to the formation of a stable dendritic array with a spacing in the range $\Lambda_{\text{min}}<\Lambda<\Lambda_{\text{max}}$. We defined new perturbation parameters, $\mathcal{A}_1=12.1905~W_0=\mathcal{A}_2=12.1905~W_0$ and $n=6$ for the initial condition. A non-zero value for $\mathcal{A}_2$ results in a superposition of two cosine waves of different wavelengths $L_x/3$ ($\kappa_1=6\pi/L_x$) and $L_x$ ($\kappa_2=2\pi/L_x$). This initial condition prevents the formation of an artificially symmetric pattern and yields a larger stable primary dendrite spacing by promoting corner-dominant growth with cell competition and elimination~\cite{RN320}. As shown in Figure~\ref{fig5}, at time $t=424~\tau_0\approx 165.41~\mathrm{s}$, the interface profile along the dendrite tips is nearly identical for both codes. While reaching full steady state would require $\approx 250~\mathrm{s}$, the comparison time is sufficient for the perturbed planar interface to develop into a nearly stable dendritic array with well-defined primary spacing. This spacing, $\Lambda\approx 200~\mathrm{\mu m}$, falls within the predicted stability band~\cite{RN163} given above, leading to a more physically consistent benchmark.

%add_revision
To provide a quantitative error estimate for this benchmark case, we extract the dendrite tip radius, $\rho$, using two independent fitting methods following Ref.~\cite{RN335}: a cross-sectional area fit and a parabolic fit of the longitudinal interface profile in the plane $x=L_x/2$, which passes through the tip of the central dendrite. Figure~\ref{fig5}(c) shows the longitudinal 2D tip profiles in the $y-z$ plane at $x=L_x/2$ for GPU-PF and PRISMS-PF, demonstrating excellent overlap near the tip. Figure~\ref{fig5}(d) shows the local tip radius obtained from the two fitting methods as a function of distance from the tip, $z_0-z$, for each code, where $z_0$ is the position of the tip along the $z$ axis. The tip radius is determined at the crossover point of the two fitted curves, yielding $\rho=13.30~W_0$ for GPU-PF and $\rho=13.87~W_0$ for PRISMS-PF, corresponding to a difference of $4.3\%$. 

To further quantify the agreement between the two tip profiles, we compute the root mean square difference (RMSD) between the two longitudinal profiles 
\begin{equation}
RMSD = \sqrt{\frac{1}{M}\sum_{i=1}^{M}(z_i^{\mathrm{GPU}} - z_i^{\mathrm{PRISMS}})^2},
\end{equation}
where the sum contains $M=113$ values for $z$ taken within one tip radius, $\rho$, from the dendrite tip, i.e. $0 \leq z_0-z \leq \rho$. Within this region, the dendrite shape is approximately axisymmetric and, therefore, the 2D profile is representative of the full 3D geometry. Beyond this region, 3D anisotropic effects and side-branch development render a 2D profile comparison ill-posed. We obtain $RMSD=0.63~W_0$, which is smaller than a single grid spacing ($\Delta x/W_0=1.2$), and is approximately $4.6\%$ of the mean tip radius. These results confirm that the two codes agree quantitatively at the tip scale to within the expected level of numerical resolution error.
%add_revision

In summary, the comparisons between the PRISMS-PF and GPU-PF codes and between different resolutions discussed above have highlighted the importance of the choice of the initial condition for the simulations. Initial planar interfaces with perturbations of different wavelengths can lead to qualitatively different dynamical behaviors ~\cite{RN318,RN319,RN320}. In some cases, the dynamical evolution can become extremely sensitive to the parameters that define the initial perturbation, which is a signature of chaotic dynamics. In such cases, the codes cannot be meaningfully compared after sufficiently long simulation time because the numerical solutions do not converge with respect to grid spacing or time step. Although the comparisons were performed for 3D systems, it should be emphasized that the issues illustrated can also be present for 2D systems.

\subsection{Performance Comparison}

\begin{figure*}[t]
\centering
\includegraphics[width=1.0\textwidth]{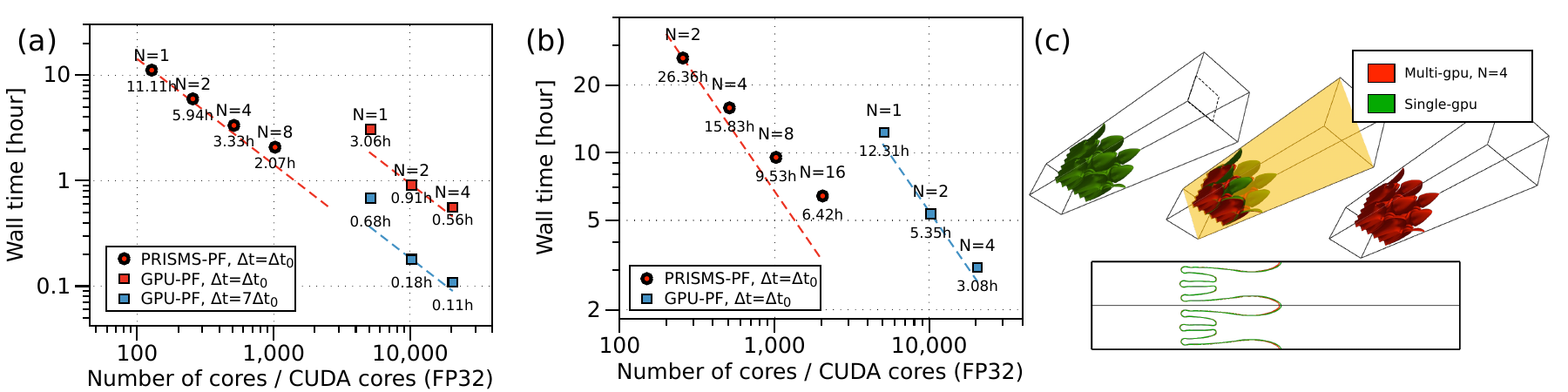}
\caption{Performance comparison between GPU-PF and PRISMS-PF.
(a) Runtime for a quarter-domain simulation with symmetry-reflected boundaries. Here $N$ denotes the number of GPUs for GPU-PF and the number of compute nodes for PRISMS-PF. Dashed lines indicate ideal strong scaling (slope $-1$).  For PRISMS-PF, the line is anchored at the smallest core count; for GPU-PF, the line is anchored at $N=2$ as the $N=1$ single-GPU run uses a fundamentally different code structure from the multi-GPU implementation and is shown only as a reference. At $N=2$ and $N=4$, improved I/O overlap yields to better GPU utilization. PRISMS-PF scales nearly ideally up to $N=2$ (256 cores) and sub-optimally for $N\ge3$, while GPU-PF scales sub-optimally (slope $\approx -0.7$) due to I/O bottlenecks.  
(b) Runtime for a full 3D domain. GPU-PF shows improved scaling (slope $\approx -0.8$) for $N=2$, and 4 GPUs, indicating efficient kernel execution and memory access. PRISMS-PF shows subideal scaling due to mesh adaptivity and communication overhead. Both codes produce nearly identical final dendritic morphologies.
(c) Final solid--liquid interface from GPU-PF comparing a quarter-domain run using no-flux mirror boundary conditions (single GPU, green) to a full-domain run using $N=4$ GPUs (black). The dashed outline shows the quarter-domain region. Both yield nearly identical morphologies, confirming convergence between the symmetry-reduced and multi-GPU implementations.}
\label{fig6}
\end{figure*}

We benchmarked GPU-PF and PRISMS-PF across two representative 3D simulation setups: (a) a reduced system occupying one-quarter of the full domain with symmetry-based reflection (Figure~\ref{fig6}a), and (b) the full 3D domain without symmetry (Figure~\ref{fig6}b). Despite the size difference, both configurations produced virtually identical final dendritic morphologies (Figure~\ref{fig6}c), validating the use of reduced-domain symmetry in GPU-PF for small-scale convergence testing.

All GPU-PF runs were executed on NVIDIA V100-SXM2 GPUs (80 SMs per GPU, 64 CUDA cores per SM), connected via NVLink (SXM2 architecture) on the Discovery cluster at Northeastern University~\cite{northeastern_rc}. Multi-GPU execution used peer-to-peer (P2P) memory access and unified memory for efficient data sharing and load distribution. 
For the reduced (quarter) domain, GPU-PF used a large timestep of $\Delta t = 7\Delta t_0$, enabled by the preconditioned PF~\cite{RN215}, completing in $0.68$~h on a single GPU, and $0.18$~h on two GPUs. 
PRISMS-PF was run on the Anvil cluster at Purdue University~\cite{Anvir}, with each node containing two AMD EPYC 7763 CPUs (64 cores per CPU, 128 cores total per node, with 256 hardware threads via simultaneous multithreading) at 2.45~GHz. 
At $\Delta t = \Delta t_0$, PRISMS-PF required $2.07$~h on 1024 cores. In this regime, GPU-PF exhibited sublinear multi-GPU scaling, whereas PRISMS-PF achieved near-ideal scaling up to 256 cores and subideal scaling for larger number of cores.

The performance benchmark for the (a) reduced system and (b) full domain all used the composite (two-wavelength) perturbations defined in Eq.~\ref{eq:perturbation}, with the same parameters used previously ($\mathcal{A}_1=12.1905~W_0=\mathcal{A}_2=12.1905~W_0$ and $n=6$). The spatial and temporal resolutions are identical to those used in Fig.~\ref{fig5}, where $\Delta x/W_0=1.2$ $\Delta t/\tau_0=5\times10^{-4}$. GPU-PF showed slightly improved multi-GPU scaling with wall times of $5.35$~h (2 GPUs), and $3.08$~h (4 GPUs). PRISMS-PF required $26.36$~h on 256 cores, $9.53$~h on 1024 cores and $6.42$~h on 2048 cores. The scaling is subideal, reflecting higher complexity and computational demand.

These results underscore the architectural tradeoffs: GPU-PF’s finite-difference implementation with preconditioning supports large stable time steps and excellent scaling on GPUs, making it well-suited for large-domain high-throughput studies. 
Conversely, PRISMS-PF’s matrix-free FEM with adaptive meshing excels at resolving sharp features on CPU-based systems, including cases with interface misorientation relative to the simulation grid, but incurs greater parallel overhead at very high core counts. 
Together, they offer complementary strategies for PF simulations in ICME workflows: GPU-PF providing peak throughput, and PRISMS-PF offering flexibility and mesh-adaptive resolution efficiency.

\section{Conclusion}
\label{sec:conclusion}
We have presented a benchmark problem for directional solidification in binary alloys, comparing two distinct numerical implementations, GPU-PF and PRISMS-PF, under a unified thin-interface PF formulation. By enforcing identical physical models, boundary conditions, and initialization protocols, we systematically isolate numerical effects to assess convergence behavior, stability, predictive accuracy, and performance across a broad parameter space. The central contribution of this work is a rigorous, direct (``apples-to-apples'') comparison of two fundamentally different phase-field implementations under a common quantitative model and experimentally relevant conditions.

The results demonstrate excellent agreement between the two codes in 2D and 3D simulations for both high-velocity solidification conditions (Section \ref{sec:2DAlcu}) and microgravity solidification regimes(Sections \ref{sec:2DDSIR} and \ref{sec:3DDSIR}). Key morphological metrics, such as the dendrite tip position, primary spacing, and interface morphology, show near-perfect overlap when the composite perturbation is used to initialize the interface. This validates both solvers under experimentally relevant conditions, where the measured spacing falls within the theoretically predicted stability range. The comparisons also highlighted the importance of the choice of initial condition in the simulations, since certain initial conditions can lead to chaotic dynamics, where the codes become incomparable.

The GPU-PF code, based on explicit finite differences with preconditioned phase fields, achieves high throughput and nearly ideal scaling for large domains on multi-GPU architectures. In contrast, PRISMS-PF, leveraging adaptive mesh refinement and matrix-free finite elements, offers greater efficiency for moderate domain sizes, better resolution near moving interfaces, and more robust convergence in cases with interface misorientation relative to the simulation grid. These complementary strengths underscore trade-offs inherent in code architecture and discretization strategies, key considerations for scaling PF simulations in ICME workflows.

This benchmark provides not only a quantitative comparison of code performance and accuracy but also a validated reference problem for the broader community. As PF modeling advances toward integration with machine learning, experimental pipelines, and materials design applications, standardized benchmarks such as those presented here are essential for reproducibility, model validation, and code verification. We encourage further development of such benchmarks, including extensions to ternary systems, polycrystalline evolution, and multi-physics couplings, to accelerate the adoption of PF modeling in engineering practice.

\section*{Data Availability}
The data that support the findings of this study is publicly available in the Materials Commons~\cite{RN312} repository: \url{https://doi.org/10.13011/m3-1rk7-k068}. DECLIC-DSI-R data used for validation is available in the NASA Physical Sciences Informatics~\cite{RN331}database.

\section*{Code Availability}
PRISMS-PF is an open-source computer code under the GNU Lesser General Public License version 2.1. The source code for PRISMS-PF is available at the following hyperlink: \url{https://github.com/prisms-center/phaseField}. The 2D and 3D finite difference codes (GPU-PF) used are available at the following hyperlink: \url{https://github.com/circs/GPU-PF/}. 

\section*{Acknowledgements}
This research was supported by the National Aeronautics and Space Administration (NASA) under award number 80NSSC24K0466, as part of the project titled \textit{Computational Modeling of Columnar-Equiaxed Alloy Solidification MicroStructures (COMPASS)}. The project is administered through the Science Mission Directorate: Biological and Physical Sciences, with funding to Northeastern University. 
% The present benchmark study was conducted as part of the NASA Biological and Physical Sciences (BPS) Integrated Computational Materials Engineering (ICME) study group effort, the results of which are documented in a companion NASA Technical Memorandum~\cite{NASA_TM_2025}. 
Additional support was provided by the U.S. Department of Energy Office of Basic Energy Sciences Division of Materials Science and Engineering under Award DE-SC0008637 as part of the Center for PRedictive Integrated Structural Materials Science (PRISMS). Computational resources were provided by the Research Computing team at Northeastern University, including access to the Discovery cluster hosting NVIDIA V100-SXM2 GPUs and our dedicated private compute node. In addition, this work used the Anvil supercomputer at the Rosen Center for Advanced Computing in Purdue University through allocation MSS160003 from the Advanced Cyberinfrastructure Coordination Ecosystem: Services \& Support (ACCESS) program~\cite{RN225}, which is supported by U.S. National Science Foundation grants \#2138259, \#2138286, \#2138307, \#2137603, and \#2138296. K.J. acknowledges partial support from Lawrence Livermore National Laboratory under Contract DE-AC52–07NA27344. 

\bibliographystyle{elsarticle-num}
\bibliography{ref.bib}

@misc{northeastern_rc,
  author       = {{Research Computing, Northeastern University}},
  title        = {Research Computing at Northeastern University},
  howpublished = {\url{https://rc.northeastern.edu/}},
  note         = {Accessed: 2025-08-11}
}

@inproceedings{Anvir,
author = {Song, X. Carol and Smith, Preston and Kalyanam, Rajesh and Zhu, Xiao and Adams, Eric and Colby, Kevin and Finnegan, Patrick and Gough, Erik and Hillery, Elizabett and Irvine, Rick and Maji, Amiya and St. John, Jason},
title = {Anvil - System Architecture and Experiences from Deployment and Early User Operations},
year = {2022},
isbn = {9781450391610},
publisher = {Association for Computing Machinery},
address = {New York, NY, USA},
doi = {10.1145/3491418.3530766},
booktitle = {Practice and Experience in Advanced Research Computing 2022: Revolutionary: Computing, Connections, You},
articleno = {23},
numpages = {9},
location = {Boston, MA, USA},
series = {PEARC '22}
}

@techreport{Lvovich2018,
  author       = {Vadim Lvovich and John Lawson},
  title        = {Integrated Computational-Experimental Development of Lithium-Air Batteries for Electric Aircraft},
  institution  = {NASA Glenn Research Center},
  year         = {2018},
  number       = {GRC-E-DAA-TN60215},
  url          = {https://ntrs.nasa.gov/citations/20190000487}
}

@article{wheeler2019pfhub,
  title = {PFHub: The Phase-Field Community Hub},
  author = {Wheeler, David and Keller, Thomas and DeWitt, Stephen J. and Jokisaari, Andrea M. and Schwen, Daniel and Guyer, Jonathan E. and Aagesen, Larry K. and Heinonen, Olle G. and Tonks, Michael R. and Voorhees, Peter W. and Warren, James A.},
  journal = {Journal of Open Research Software},
  volume = {7},
  number = {1},
  pages = {29},
  year = {2019},
  doi = {10.5334/jors.276},
}

@techreport{osti_1729722,
  author = {Dingreville, Remi Philippe Michel and Stewart, James Allen and Chen, Elton Y. and Monti, Joseph Michael},
  title = {Benchmark problems for the Mesoscale Multiphysics Phase Field Simulator (MEMPHIS)},
  institution = {Sandia National Laboratories},
  year = {2020},
  doi = {10.2172/1729722},
}

@article{giudicelli2024moose,
   title = {3.0 - {MOOSE}: Enabling massively parallel multiphysics simulations},
   author = {Guillaume Giudicelli and Alexander Lindsay and Logan Harbour and Casey Icenhour and
             Mengnan Li and Joshua E. Hansel and Peter German and Patrick Behne and Oana Marin and
             Roy H. Stogner and Jason M. Miller and Daniel Schwen and Yaqi Wang and Lynn Munday and
             Sebastian Schunert and Benjamin W. Spencer and Dewen Yushu and Antonio Recuero and
             Zachary M. Prince and Max Nezdyur and Tianchen Hu and Yinbin Miao and
             Yeon Sang Jung and Christopher Matthews and April Novak and Brandon Langley and
             Timothy Truster and Nuno Nobre and Brian Alger and David Andr{\v{s}} and
             Fande Kong and Robert Carlsen and Andrew E. Slaughter and John W. Peterson and
             Derek Gaston and Cody Permann},
    year = {2024},
 journal = {{SoftwareX}},
  volume = {26},
   pages = {101690},
    issn = {2352-7110},
     doi = {10.1016/j.softx.2024.101690},
}

@incollection{Fix1983,
  author = {G. J. Fix},
  title = {Free Boundary Problems: Theory and Applications},
  booktitle = {Free Boundary Problems: Theory and Applications},
  editor = {A. Fasano and M. Primicerio},
  pages = {580},
  publisher = {Pitman},
  year = {1983},
  address = {Boston},
  type = {Book Section}
}

@book{boettinger2010phasefield,
  title     = {Phase‐Field Methods in Materials Science and Engineering},
  year      = {2010},
  publisher = {Wiley-VCH},
  doi       = {10.1002/9783527631520},
  editor    = {Boettinger, William J. and Warren, James A. and Beckermann, Christoph and Karma, Alain}
}

@article{RN2,
   author = {Chen, C. H. and Cambonie, T. and Lazarus, V. and Nicoli, M. and Pons, A. J. and Karma, A.},
   title = {Crack Front Segmentation and Facet Coarsening in Mixed-Mode Fracture},
   journal = {Phys Rev Lett},
   volume = {115},
   number = {26},
   pages = {265503},
   note = {Chen, Chih-Hung
Cambonie, Tristan
Lazarus, Veronique
Nicoli, Matteo
Pons, Antonio J
Karma, Alain
eng
Phys Rev Lett. 2015 Dec 31;115(26):265503. doi: 10.1103/PhysRevLett.115.265503. Epub 2015 Dec 30.},
   ISSN = {1079-7114 (Electronic)
0031-9007 (Linking)},
   DOI = {10.1103/PhysRevLett.115.265503},
   year = {2015},
   type = {Journal Article}
}

@article{RN14,
   author = {Chen, C. H. and Bouchbinder, E. and Karma, A.},
   title = {Instability in dynamic fracture and the failure of the classical theory of cracks},
   journal = {Nature Physics},
   volume = {13},
   number = {12},
   pages = {1186-+},
   ISSN = {1745-2473},
   DOI = {10.1038/Nphys4237},
   year = {2017},
   type = {Journal Article}
}

@article{RN32,
  title = {Quantitative phase-field model of alloy solidification},
  author = {Echebarria, Blas and Folch, Roger and Karma, Alain and Plapp, Mathis},
  journal = {Phys. Rev. E},
  volume = {70},
  issue = {6},
  pages = {061604},
  numpages = {22},
  year = {2004},
  month = {Dec},
  publisher = {American Physical Society},
  doi = {10.1103/PhysRevE.70.061604},
type = {Journal Article}
}

@article{RN34,
   author = {Karma, Alain and Rappel, Wouter-Jan},
   title = {Quantitative phase-field modeling of dendritic growth in two and three dimensions},
   journal = {Physical Review E},
   volume = {57},
   number = {4},
   pages = {4323-4349},
   note = {PRE},
   DOI = {10.1103/PhysRevE.57.4323},
   year = {1998},
   type = {Journal Article}
}

@article{RN38,
   author = {Ji, Kaihua and Tabrizi, Amirhossein Molavi and Karma, Alain},
   title = {Isotropic finite-difference approximations for phase-field simulations of polycrystalline alloy solidification},
   journal = {Journal of Computational Physics},
   volume = {457},
   pages = {111069},
   ISSN = {0021-9991},
   DOI = {https://doi.org/10.1016/j.jcp.2022.111069},
   year = {2022},
   type = {Journal Article}
}

@article{RN43,
   author = {Song, Y. and Tourret, D. and Mota, F. L. and Pereda, J. and Billia, B. and Bergeon, N. and Trivedi, R. and Karma, A.},
   title = {Thermal-field effects on interface dynamics and microstructure selection during alloy directional solidification},
   journal = {Acta Materialia},
   volume = {150},
   pages = {139-152},
   ISSN = {1359-6454},
   DOI = {10.1016/j.actamat.2018.03.012},
   year = {2018},
   type = {Journal Article}
}

@article{RN45,
   author = {Clarke, A. J. and Tourret, D. and Song, Y. and Imhoff, S. D. and Gibbs, P. J. and Gibbs, J. W. and Fezzaa, K. and Karma, A.},
   title = {Microstructure selection in thin-sample directional solidification of an Al-Cu alloy: In situ X-ray imaging and phase-field simulations},
   journal = {Acta Materialia},
   volume = {129},
   pages = {203-216},
   ISSN = {1359-6454},
   DOI = {https://doi.org/10.1016/j.actamat.2017.02.047},
   year = {2017},
   type = {Journal Article}
}

@article{RN53,
   author = {Tourret, D. and Karma, A.},
   title = {Multiscale dendritic needle network model of alloy solidification},
   journal = {Acta Materialia},
   volume = {61},
   number = {17},
   pages = {6474-6491},
   ISSN = {1359-6454},
   DOI = {https://doi.org/10.1016/j.actamat.2013.07.026},
   year = {2013},
   type = {Journal Article}
}

@article{song2023cell,
  title={Cell invasion during competitive growth of polycrystalline solidification patterns},
  author={Song, Younggil and Mota, Fatima L and Tourret, Damien and Ji, Kaihua and Billia, Bernard and Trivedi, Rohit and Bergeon, Nathalie and Karma, Alain},
  journal={Nature Communications},
  volume={14},
  number={1},
  pages={2244},
  year={2023},
  publisher={Nature Publishing Group UK London}
}

@article{mota2023influence,
  title={Influence of macroscopic interface curvature on dendritic patterns during directional solidification of bulk samples: experimental and phase-field studies},
  author={Mota, FL and Ji, K and Littles, L Strutzenberg and Trivedi, R and Karma, A and Bergeon, N},
  journal={Acta Materialia},
  volume={250},
  pages={118849},
  year={2023},
  publisher={Elsevier}
}

@article{RN59,
   author = {Mota, F. L. and Bergeon, N. and Tourret, D. and Karma, A. and Trivedi, R. and Billia, B.},
   title = {Initial transient behavior in directional solidification of a bulk transparent model alloy in a cylinder},
   journal = {Acta Materialia},
   volume = {85},
   pages = {362-377},
   ISSN = {1359-6454},
   DOI = {10.1016/j.actamat.2014.11.024},
   year = {2015},
   type = {Journal Article}
}

@article{RN89,
   author = {Echebarria, Blas and Karma, Alain and Gurevich, Sebastian},
   title = {Onset of sidebranching in directional solidification},
   journal = {Physical Review E},
   volume = {81},
   number = {2},
   pages = {021608},
   note = {PRE},
   DOI = {10.1103/PhysRevE.81.021608},
   year = {2010},
   type = {Journal Article}
}

@inbook{RN91,
   author = {Langer, J. S.},
   title = {MODELS OF PATTERN FORMATION IN FIRST-ORDER PHASE TRANSITIONS},
   booktitle = {Directions in Condensed Matter Physics},
   publisher = {World Scientific Publishing Co. Pte. Ltd.},
   pages = {165--186},
   DOI = {10.1142/9789814415309_0005},
   year = {1986},
   type = {Book Section}
}

@article{RN92,
   author = {Kim, Seong Gyoon and Kim, Won Tae and Suzuki, Toshio},
   title = {Phase-field model for binary alloys},
   journal = {Physical Review E},
   volume = {60},
   number = {6},
   pages = {7186-7197},
   note = {PRE},
   DOI = {10.1103/PhysRevE.60.7186},
   year = {1999},
   type = {Journal Article}
}

@article{RN98,
   author = {DeWitt, Stephen and Rudraraju, Shiva and Montiel, David and Andrews, W. Beck and Thornton, Katsuyo},
   title = {PRISMS-PF: A general framework for phase-field modeling with a matrix-free finite element method},
   journal = {npj Computational Materials},
   volume = {6},
   number = {1},
   pages = {29},
   ISSN = {2057-3960},
   DOI = {10.1038/s41524-020-0298-5},
   year = {2020},
   type = {Journal Article}
}

@article{RN111,
   author = {Bergeon, N. and Trivedi, R. and Billia, B. and Echebarria, B. and Karma, A. and Liu, S. and Weiss, C. and Mangelinck, N.},
   title = {Necessity of investigating microstructure formation during directional solidification of transparent alloys in 3D},
   journal = {Advances in Space Research},
   volume = {36},
   number = {1},
   pages = {80-85},
   ISSN = {0273-1177},
   DOI = {10.1016/j.asr.2005.02.092},
   year = {2005},
   type = {Journal Article}
}

@article{RN151,
   author = {Altmann, S. L. and Cracknell, A. P.},
   title = {Lattice Harmonics I. Cubic Groups},
   journal = {Reviews of Modern Physics},
   volume = {37},
   number = {1},
   pages = {19-32},
   note = {RMP},
   DOI = {10.1103/RevModPhys.37.19},
   url = {https://link.aps.org/doi/10.1103/RevModPhys.37.19},
   year = {1965},
   type = {Journal Article}
}

@article{RN163,
   author = {Medjkoune, Mehdi and Lyons, Trevor and Mota, Fátima L. and Tian, Jiefu and Ji, Kaihua and Littles, Louise and Karma, Alain and Bergeon, Nathalie},
   title = {Benchmark microgravity experiments and computations for 3D dendritic-array stability in directional solidification},
   journal = {Acta Materialia},
   volume = {292},
   pages = {120954},
   ISSN = {1359-6454},
   DOI = {10.1016/j.actamat.2025.120954},
   year = {2025},
   type = {Journal Article}
}

@article{RN164,
   author = {Collins, Joseph B. and Levine, Herbert},
   title = {Diffuse interface model of diffusion-limited crystal growth},
   journal = {Physical Review B},
   volume = {31},
   number = {9},
   pages = {6119-6122},
   note = {PRB},
   DOI = {10.1103/PhysRevB.31.6119},
   year = {1985},
   type = {Journal Article}
}

@article{RN165,
   author = {Cahn, John W. and Hilliard, John E.},
   title = {Free Energy of a Nonuniform System. I. Interfacial Free Energy},
   journal = {The Journal of Chemical Physics},
   volume = {28},
   number = {2},
   pages = {258-267},
   ISSN = {0021-9606},
   DOI = {10.1063/1.1744102},
   year = {1958},
   type = {Journal Article}
}

@article{RN166,
   author = {Tonks, Michael R. and Aagesen, Larry K.},
   title = {The Phase Field Method: Mesoscale Simulation Aiding Material Discovery},
   journal = {Annual Review of Materials Research},
   volume = {49},
   number = {Volume 49, 2019},
   pages = {79-102},
   ISSN = {1545-4118},
   DOI = {10.1146/annurev-matsci-070218-010151},
   year = {2019},
   type = {Journal Article}
}

@article{RN167,
   author = {Zhuang, X. and Zhou, S. and Huynh, G. D. and Areias, P. and Rabczuk, T.},
   title = {Phase field modeling and computer implementation: A review},
   journal = {Engineering Fracture Mechanics},
   volume = {262},
   pages = {108234},
   ISSN = {0013-7944},
   DOI = {10.1016/j.engfracmech.2022.108234},
   year = {2022},
   type = {Journal Article}
}

@article{RN168,
   author = {Jokisaari, A. M. and Voorhees, P. W. and Guyer, J. E. and Warren, J. A. and Heinonen, O. G.},
   title = {Phase field benchmark problems for dendritic growth and linear elasticity},
   journal = {Computational Materials Science},
   volume = {149},
   pages = {336-347},
   ISSN = {0927-0256},
   DOI = {https://doi.org/10.1016/j.commatsci.2018.03.015},
   year = {2018},
   type = {Journal Article}
}

@article{RN171,
   author = {Karma, Alain and Rappel, Wouter-Jan},
   title = {Phase-field method for computationally efficient modeling of solidification with arbitrary interface kinetics},
   journal = {Physical Review E},
   volume = {53},
   number = {4},
   pages = {R3017-R3020},
   note = {PRE},
   DOI = {10.1103/PhysRevE.53.R3017},
   year = {1996},
   type = {Journal Article}
}

@article{RN172,
   author = {Grabowski, Jeff and Sebastian, Jason and Olson, Greg and Asphahani, Aziz and Genellie, Raymond, Jr.},
   title = {Integrated Computational Materials Engineering Helps Successfully Develop Aerospace Alloys},
   journal = {AM\&P Technical Articles},
   volume = {171},
   number = {9},
   pages = {17-19},
   ISSN = {0882-7958},
   DOI = {10.31399/asm.amp.2013-09.p017},
   year = {2013},
   type = {Journal Article}
}

@article{RN173,
   author = {Horstemeyer, M. F. and Wang, P.},
   title = {Cradle-to-grave simulation-based design incorporating multiscale microstructure-property modeling: Reinvigorating design with science},
   journal = {Journal of Computer-Aided Materials Design},
   volume = {10},
   number = {1},
   pages = {13-34},
   ISSN = {1573-4900},
   DOI = {10.1023/B:JCAD.0000024171.13480.24},
   year = {2003},
   type = {Journal Article}
}

@article{RN174,
   author = {Joost, William J.},
   title = {Reducing Vehicle Weight and Improving U.S. Energy Efficiency Using Integrated Computational Materials Engineering},
   journal = {JOM},
   volume = {64},
   number = {9},
   pages = {1032-1038},
   ISSN = {1543-1851},
   DOI = {10.1007/s11837-012-0424-z},
   year = {2012},
   type = {Journal Article}
}

@article{RN177,
   author = {Gránásy, László and Börzsönyi, Tamás and Pusztai, Tamás},
   title = {Nucleation and Bulk Crystallization in Binary Phase Field Theory},
   journal = {Physical Review Letters},
   volume = {88},
   number = {20},
   pages = {206105},
   note = {PRL},
   DOI = {10.1103/PhysRevLett.88.206105},
   year = {2002},
   type = {Journal Article}
}

@article{RN178,
   author = {Karma, Alain and Plapp, Mathis},
   title = {Spiral Surface Growth without Desorption},
   journal = {Physical Review Letters},
   volume = {81},
   number = {20},
   pages = {4444-4447},
   note = {PRL},
   DOI = {10.1103/PhysRevLett.81.4444},
   year = {1998},
   type = {Journal Article}
}

@article{RN179,
   author = {Elder, K. R. and Katakowski, Mark and Haataja, Mikko and Grant, Martin},
   title = {Modeling Elasticity in Crystal Growth},
   journal = {Physical Review Letters},
   volume = {88},
   number = {24},
   pages = {245701},
   note = {PRL},
   DOI = {10.1103/PhysRevLett.88.245701},
   year = {2002},
   type = {Journal Article}
}

@article{RN180,
   author = {Moelans, Nele and Blanpain, Bart and Wollants, Patrick},
   title = {An introduction to phase-field modeling of microstructure evolution},
   journal = {Calphad},
   volume = {32},
   number = {2},
   pages = {268-294},
   ISSN = {0364-5916},
   DOI = {https://doi.org/10.1016/j.calphad.2007.11.003},
   year = {2008},
   type = {Journal Article}
}

@inbook{RN181,
   author = {Karma, Alain},
   title = {Phase-Field Modeling},
   booktitle = {Handbook of Materials Modeling: Methods},
   publisher = {Springer Netherlands},
   address = {Dordrecht},
   pages = {2087--2103},
   ISBN = {978-1-4020-3286-8},
   DOI = {10.1007/978-1-4020-3286-8_108},
   year = {2005},
   type = {Book Section}
}

@article{RN182,
   author = {Chen, Long-Qing},
   title = {Phase-Field Models for Microstructure Evolution},
   journal = {Annual Review of Materials Research},
   volume = {32},
   number = {Volume 32, 2002},
   pages = {113-140},
   ISSN = {1545-4118},
   DOI = {https://doi.org/10.1146/annurev.matsci.32.112001.132041},
   year = {2002},
   type = {Journal Article}
}

@article{RN184,
   author = {Schwen, D. and Aagesen, L. K. and Peterson, J. W. and Tonks, M. R.},
   title = {Rapid multiphase-field model development using a modular free energy based approach with automatic differentiation in MOOSE/MARMOT},
   journal = {Computational Materials Science},
   volume = {132},
   pages = {36-45},
   ISSN = {0927-0256},
   DOI = {https://doi.org/10.1016/j.commatsci.2017.02.017},
   year = {2017},
   type = {Journal Article}
}

@article{RN185,
   author = {Tonks, Michael R. and Gaston, Derek and Millett, Paul C. and Andrs, David and Talbot, Paul},
   title = {An object-oriented finite element framework for multiphysics phase field simulations},
   journal = {Computational Materials Science},
   volume = {51},
   number = {1},
   pages = {20-29},
   ISSN = {0927-0256},
   DOI = {https://doi.org/10.1016/j.commatsci.2011.07.028},
   year = {2012},
   type = {Journal Article}
}

@inbook{RN186,
   author = {DeWitt, Stephen and Thornton, Katsuyo},
   title = {Phase Field Modeling of Microstructural Evolution},
   booktitle = {Computational Materials System Design},
   publisher = {Springer International Publishing},
   address = {Cham},
   pages = {67--87},
   ISBN = {978-3-319-68280-8},
   DOI = {10.1007/978-3-319-68280-8_4},
   year = {2018},
   type = {Book Section}
}

@article{RN187,
   author = {Turnali, Ahmet and Kibaroglu, Dilay and Evers, Nico and Gehlmann, Jaqueline and Sayk, Lennart and Peter, Nicolas J. and Elsayed, Abdelrahman and Noori, Mehdi and Allam, Tarek and Schleifenbaum, Johannes Henrich and Haase, Christian},
   title = {Segregation-guided alloy design via tailored solidification behavior},
   journal = {Materials Today Advances},
   volume = {25},
   pages = {100549},
   ISSN = {2590-0498},
   DOI = {https://doi.org/10.1016/j.mtadv.2024.100549},
   year = {2025},
   type = {Journal Article}
}

@article{RN188,
   author = {Boussinot, Guillaume and Cazic, Ivan and Döring, Markus and Schmidt, Michael and Apel, Markus},
   title = {Stabilization of the Ternary Eutectic Growth in Additively Manufactured Al-Ni-Ce Alloys},
   journal = {Advanced Engineering Materials},
   volume = {n/a},
   number = {n/a},
   pages = {2401665},
   ISSN = {1438-1656},
   DOI = {https://doi.org/10.1002/adem.202401665},
   year = {2024},
   type = {Journal Article}
}

@article{RN189,
   author = {Guyer, Jonathan E. and Wheeler, Daniel and Warren, James A.},
   title = {FiPy: Partial Differential Equations with Python},
   journal = {Computing in Science \& Engineering},
   volume = {11},
   number = {3},
   pages = {6-15},
   year = {2009},
   type = {Journal Article}
}

@article{RN190,
   author = {Mohanty, R. R. and Guyer, J. E. and Sohn, Y. H.},
   title = {Diffusion under temperature gradient: A phase-field model study},
   journal = {Journal of Applied Physics},
   volume = {106},
   number = {3},
   ISSN = {0021-8979},
   DOI = {10.1063/1.3190607},
   year = {2009},
   type = {Journal Article}
}

@article{RN191,
   author = {Cheng, Yongfu and Wang, Gang and Qiu, Zhaoguo and Zheng, Zhigang and Zeng, Dechang and Tang, Xu and Shi, Rongpei and Uddagiri, Murali and Steinbach, Ingo},
   title = {Multi-physics simulation of non-equilibrium solidification in Ti-Nb alloy during selective laser melting},
   journal = {Acta Materialia},
   volume = {272},
   pages = {119923},
   ISSN = {1359-6454},
   DOI = {https://doi.org/10.1016/j.actamat.2024.119923},
   year = {2024},
   type = {Journal Article}
}

@article{RN192,
   author = {Tegeler, Marvin and Shchyglo, Oleg and Kamachali, Reza Darvishi and Monas, Alexander and Steinbach, Ingo and Sutmann, Godehard},
   title = {Parallel multiphase field simulations with OpenPhase},
   journal = {Computer physics communications},
   volume = {215},
   pages = {173-187},
   ISSN = {0010-4655},
   year = {2017},
   type = {Journal Article}
}

@article{RN194,
   author = {Plapp, Mathis and Karma, Alain},
   title = {Multiscale Finite-Difference-Diffusion-Monte-Carlo Method for Simulating Dendritic Solidification},
   journal = {Journal of Computational Physics},
   volume = {165},
   number = {2},
   pages = {592-619},
   ISSN = {0021-9991},
   DOI = {https://doi.org/10.1006/jcph.2000.6634},
   year = {2000},
   type = {Journal Article}
}

@article{RN210,
   author = {Greenwood, Michael and Shampur, K. N. and Ofori-Opoku, Nana and Pinomaa, Tatu and Wang, Lei and Gurevich, Sebastian and Provatas, Nikolas},
   title = {Quantitative 3D phase field modelling of solidification using next-generation adaptive mesh refinement},
   journal = {Computational Materials Science},
   volume = {142},
   pages = {153-171},
   ISSN = {0927-0256},
   DOI = {https://doi.org/10.1016/j.commatsci.2017.09.029},
   year = {2018},
   type = {Journal Article}
}

@article{RN211,
   author = {Plapp, Mathis and Karma, Alain},
   title = {Multiscale Random-Walk Algorithm for Simulating Interfacial Pattern Formation},
   journal = {Physical Review Letters},
   volume = {84},
   number = {8},
   pages = {1740-1743},
   note = {PRL},
   DOI = {10.1103/PhysRevLett.84.1740},
   year = {2000},
   type = {Journal Article}
}

@article{RN212,
   author = {Provatas, Nikolas and Goldenfeld, Nigel and Dantzig, Jonathan},
   title = {Efficient Computation of Dendritic Microstructures Using Adaptive Mesh Refinement},
   journal = {Physical Review Letters},
   volume = {80},
   number = {15},
   pages = {3308-3311},
   note = {PRL},
   DOI = {10.1103/PhysRevLett.80.3308},
   year = {1998},
   type = {Journal Article}
}

@article{karma2001phase,
  title={Phase-field formulation for quantitative modeling of alloy solidification},
  author={Karma, Alain},
  journal={Physical review letters},
  volume={87},
  number={11},
  pages={115701},
  year={2001},
  publisher={APS}
}

@article{RN214,
   author = {Hötzer, J. and Reiter, A. and Hierl, H. and Steinmetz, P. and Selzer, M. and Nestler, Britta},
   title = {The parallel multi-physics phase-field framework Pace3D},
   journal = {Journal of Computational Science},
   volume = {26},
   pages = {1-12},
   ISSN = {1877-7503},
   DOI = {https://doi.org/10.1016/j.jocs.2018.02.011},
   year = {2018},
   type = {Journal Article}
}

@article{RN215,
   author = {Glasner, Karl},
   title = {Nonlinear Preconditioning for Diffuse Interfaces},
   journal = {Journal of Computational Physics},
   volume = {174},
   number = {2},
   pages = {695-711},
   ISSN = {0021-9991},
   DOI = {https://doi.org/10.1006/jcph.2001.6933},
   year = {2001},
   type = {Journal Article}
}

@article{RN217,
   author = {Plapp, Mathis},
   title = {Unified derivation of phase-field models for alloy solidification from a grand-potential functional},
   journal = {Physical Review E},
   volume = {84},
   number = {3},
   pages = {031601},
   note = {PRE},
   DOI = {10.1103/PhysRevE.84.031601},
   year = {2011},
   type = {Journal Article}
}

@article{RN220,
   author = {Allison, John and Backman, Dan and Christodoulou, Leo},
   title = {Integrated computational materials engineering: A new paradigm for the global materials profession},
   journal = {JOM},
   volume = {58},
   number = {11},
   pages = {25-27},
   ISSN = {1543-1851},
   DOI = {10.1007/s11837-006-0223-5},
   year = {2006},
   type = {Journal Article}
}

@article{RN224,
    url = {https://doi.org/10.1515/jnma-2023-0089},
    title = {The deal.II Library, Version 9.5},
    author = {Daniel Arndt and Wolfgang Bangerth and Maximilian Bergbauer and Marco Feder and Marc Fehling and Johannes Heinz and Timo Heister and Luca Heltai and Martin Kronbichler and Matthias Maier and Peter Munch and Jean-Paul Pelteret and Bruno Turcksin and David Wells and Stefano Zampini},
    pages = {231--246},
    volume = {31},
    number = {3},
    journal = {Journal of Numerical Mathematics},
    doi = {doi:10.1515/jnma-2023-0089},
    year = {2023},
    lastchecked = {2025-07-03}
}

@inproceedings{RN225,
author = {Boerner, Timothy J. and Deems, Stephen and Furlani, Thomas R. and Knuth, Shelley L. and Towns, John},
title = {ACCESS: Advancing Innovation: NSF’s Advanced Cyberinfrastructure Coordination Ecosystem: Services \& Support},
year = {2023},
isbn = {9781450399852},
publisher = {Association for Computing Machinery},
address = {New York, NY, USA},
url = {https://doi.org/10.1145/3569951.3597559},
doi = {10.1145/3569951.3597559},
booktitle = {Practice and Experience in Advanced Research Computing 2023: Computing for the Common Good},
pages = {173–176},
numpages = {4},
location = {Portland, OR, USA},
series = {PEARC '23}
}

@article{RN311,
   author = {Yao, Zhenjie and Montiel, David and Allison, John},
   title = {Investigating the Effects of Dendrite Evolution on Microsegregation in Al–Cu Alloys by Coupling Experiments, Micro-modeling, and Phase-Field Simulations},
   journal = {Metallurgical and Materials Transactions A},
   volume = {53},
   number = {9},
   pages = {3341-3356},
   ISSN = {1543-1940},
   DOI = {10.1007/s11661-022-06748-5},
   url = {https://doi.org/10.1007/s11661-022-06748-5},
   year = {2022},
   type = {Journal Article}
}

@article{RN312,
   author = {Puchala, Brian and Tarcea, Glenn and Marquis, Emmanuelle A. and Hedstrom, Margaret and Jagadish, H. V. and Allison, John E.},
   title = {The Materials Commons: A Collaboration Platform and Information Repository for the Global Materials Community},
   journal = {JOM},
   volume = {68},
   number = {8},
   pages = {2035-2044},
   ISSN = {1543-1851},
   DOI = {10.1007/s11837-016-1998-7},
   url = {https://doi.org/10.1007/s11837-016-1998-7},
   year = {2016},
   type = {Journal Article}
}

@article{RN318,
   author = {Mullins, W. W. and Sekerka, R. F.},
   title = {Morphological Stability of a Particle Growing by Diffusion or Heat Flow},
   journal = {Journal of Applied Physics},
   volume = {34},
   number = {2},
   pages = {323-329},
   ISSN = {0021-8979},
   DOI = {10.1063/1.1702607},
   year = {1963},
   type = {Journal Article}
}

@article{RN319,
   author = {Mullins, W. W. and Sekerka, R. F.},
   title = {Stability of a Planar Interface During Solidification of a Dilute Binary Alloy},
   journal = {Journal of Applied Physics},
   volume = {35},
   number = {2},
   pages = {444-451},
   ISSN = {0021-8979},
   DOI = {10.1063/1.1713333},
   year = {1964},
   type = {Journal Article}
}

@article{RN320,
   author = {Langer, J. S.},
   title = {Instabilities and pattern formation in crystal growth},
   journal = {Reviews of Modern Physics},
   volume = {52},
   number = {1},
   pages = {1-28},
   note = {RMP},
   DOI = {10.1103/RevModPhys.52.1},
   year = {1980},
   type = {Journal Article}
}

@article{RN329,
   author = {Lan, C. W. and Shih, C. J. and Lee, M. H.},
   title = {Quantitative phase field simulation of deep cells in directional solidification of an alloy},
   journal = {Acta Materialia},
   volume = {53},
   number = {8},
   pages = {2285-2294},
   ISSN = {1359-6454},
   DOI = {https://doi.org/10.1016/j.actamat.2005.01.034},
   year = {2005},
   type = {Journal Article}
}

@article{RN330,
   author = {Diepers, H. J. and Ma, D. and Steinbach, I.},
   title = {History effects during the selection of primary dendrite spacing. Comparison of phase-field simulations with experimental observations},
   journal = {Journal of Crystal Growth},
   volume = {237-239},
   pages = {149-153},
   ISSN = {0022-0248},
   DOI = {https://doi.org/10.1016/S0022-0248(01)01932-7},
   year = {2002},
   type = {Journal Article}
}

@misc{RN331,
   author = {Karma, Alain and Trivedi, Rohit and Bergeon, Nathalie and Mota, Fatima and Billia, Bernard and Littles, Louise},
   title = {{DEvice} for the study of {Critical} {LIquids} and {Crystallization} {Directional} {Solidification} {Insert}-{Reflight}({DECLIC} {DSI}-{R})},
   publisher = {NASA PSI Data Repository},
   note = {{NASA} PSI Data Repository, Version 5},
   DOI = {10.60555/wc3d-md27},
   year = {2023},
   type = {Dataset}
}

@article{RN335,
   author = {Karma, Alain and Lee, Youngyih H. and Plapp, Mathis},
   title = {Three-dimensional dendrite-tip morphology at low undercooling},
   journal = {Physical Review E},
   volume = {61},
   number = {4},
   pages = {3996-4006},
   note = {PRE},
   DOI = {10.1103/PhysRevE.61.3996},
   year = {2000},
   type = {Journal Article}
}

@techreport{NASA_TM_2025,
  author      = {O'Connor, Andrew and Littles, Louise and Richter, Brodan 
                 and Glaessgen, Edward and Karma, Alain and Matson, Douglas 
                 and Voorhees, Peter and Michael, Fredrick and Rupp, Benjamin 
                 and SanSoucie, Michael and Sowards, Jeffrey and West, Jeffrey 
                 and Weber, George R. and Pribe, Joshua D. and Yamakov, Vesselin 
                 and Yeratapally, Saikumar R. and Cole, Vernon and Waxman, Rae},
  title       = {Reduced Gravity and Microgravity Integrated Computational 
                 Materials Engineering ({ICME})},
  institution = {National Aeronautics and Space Administration},
  year        = {2025},
  month       = {April},
  number      = {NASA/TM-20250000717},
  address     = {Marshall Space Flight Center, Huntsville, AL},
  url         = {https://ntrs.nasa.gov/citations/20250000717},
}

\end{document}